\documentclass[aps,pre,superscriptaddress,floatfix,tightenlines,showpacs,twocolumn,10pt,notitlepage]{revtex4-2}
\usepackage{amsmath}
\usepackage{amssymb}
\usepackage{amsthm}
\usepackage{bm}
\usepackage{bbm}
\usepackage[shortlabels]{enumitem}
\usepackage{tabularx}
\usepackage{graphicx}
\usepackage{tikz-cd}
\usepackage{xcolor}
\usepackage{hyperref}
\usepackage[normalem]{ulem}
\usepackage[textsize=scriptsize,backgroundcolor=white]{todonotes}
\usepackage{url}
\usepackage[dvipsnames,svgnames,table]{xcolor}
\usepackage{framed}
\usepackage{color} 
\usepackage{mathtools}
\usepackage{tikz, quantikz}
\usetikzlibrary{shapes.callouts}

\hypersetup{
	colorlinks=true,
	linkcolor=blue,
	citecolor=blue,
	urlcolor=blue,
	filecolor=blue,
	anchorcolor = green,
}

\begin{document}
\title{Quantum-enhanced Markov chain Monte Carlo sampling to model Lagrangian tracer dispersion in turbulent boundary layer}

\author{Fabian Schindler}
\affiliation{Institute of Thermodynamics and Fluid Mechanics, Technische Universit\"at Ilmenau, P.O.\ Box 100565, D-98684 Ilmenau, Germany}

\author{Jörg Schumacher}
\affiliation{Institute of Thermodynamics and Fluid Mechanics, Technische Universit\"at Ilmenau, P.O.\ Box 100565, D-98684 Ilmenau, Germany}

\date{\today}

\begin{abstract}
We present a quantum-enhanced Markov chain Monte Carlo (QE-MCMC) method to sample turbulent acceleration vectors from a joint target distribution that depends on all three components and height to model the transport and dispersion of massless Lagrangian tracer particles in two different turbulent shear flows. A homogeneous shear flow, which is characterized by a uniform shear rate $S$, is considered as the starting point. Secondly, a turbulent boundary layer, which forms in both halves of a plane turbulent channel flow (TCF) at a friction Reynolds number of $Re_{\tau}=1000$, is considered. Then the mean shear rate $S(y)$ varies with distance from the wall $y$. In this hybrid quantum-classical method, the proposal distribution $Q$ for the first of the two Metropolis-Hastings sampling substeps is constructed by a parametric quantum circuit. The algorithm generates synthetic tracer particle tracks. The resulting scaling laws for the tracer particle pair dispersion, a central quantity to probe the turbulent mixing properties of the shear flow from a Lagrangian perspective, are found to agree with a stochastic transport model consisting of a set of coupled Langevin equations and the classical MCMC counterpart. Differently to the classical sampling method, QE-MCMC takes to this end a tempered target distribution. Due to the height-dependence of the tracer dynamics in TCF, an effective height-weighted spectral gap between the first and second eigenvalue of the Markov chain transition matrix is introduced. The latter is found to significantly exceed the one of classical MCMC when the sampling is done from a multivariate distribution with cross correlations at the highest qubit numbers and thus resolutions. Consequently, our results support the applicability of this one-shot algorithm as a generative Lagrangian quantum computing module, possibly embedded in a complex fluid flow problem. Our module is found to work reliably for a relatively small number of qubits per spatial dimension of $N_q\le 6$.  
\end{abstract}

\maketitle

\section{Introduction}
In many applications in science and technology, one has to sample from a complex probability density function $P(\mathbf{a})$, where the data (or state) vector $\mathbf{a} \in \mathbb{R}^d$ and $d\gg 1$. A typical example is a multivariate normal distribution with a potentially complex covariance matrix. Direct sampling may be computationally infeasible or analytically intractable. To address this challenge, Markov Chain Monte Carlo (MCMC) methods provide a powerful framework for generating approximate samples from distributions that depend on many variables \cite{Bishop2006}. Its most prominent version goes back to the pioneering works of Metropolis et al. \cite{Metropolis1953} and Hastings \cite{Hastings1970}.
The method consists basically of two steps; in a first step a new sample $\mathbf{a}^*$ is created by a proposal distribution $Q(\mathbf{a}^*|\mathbf{a})$ and in a second step this new sample is accepted or rejected following an acceptance probability $A(\mathbf{a}^*|\mathbf{a})$. The transition probability $\Pi$ for the Markov chain step from $\mathbf{a}$ to $\mathbf{a}^*$ is then given by
\begin{equation}
    \Pi(\mathbf{a}^*|\mathbf{a})=Q(\mathbf{a}^*|\mathbf{a}) A(\mathbf{a}^*|\mathbf{a}) \quad \forall\; \mathbf{a}^*\ne \mathbf{a}\,.
    \label{eq:Pi}
\end{equation}
Classical MCMC methods are widely used. In fluid mechanics and atmospheric or environmental sciences, MCMC methods are used for data assimilation, i.e., to integrate observations into fluid mechanical models, such as for water cycle processes \cite{Vrugt2013}. They are also applied to estimate parameters and uncertainties in complex flow problems, e.g., in numerical simulations of volcanic lava flows which were supported by satellite data \cite{Zucarello2025}. Classical MCMC methods have been coupled to quantum algorithms in chemistry, such as variational quantum eigensolvers \cite{Cerezo2021} to accelerate high-dimensional non-convex optimization, e.g., in graph partitioning problems \cite{Patti2023}.   

\begin{figure*}[t]
    \centering
    \includegraphics[width=0.85\textwidth]{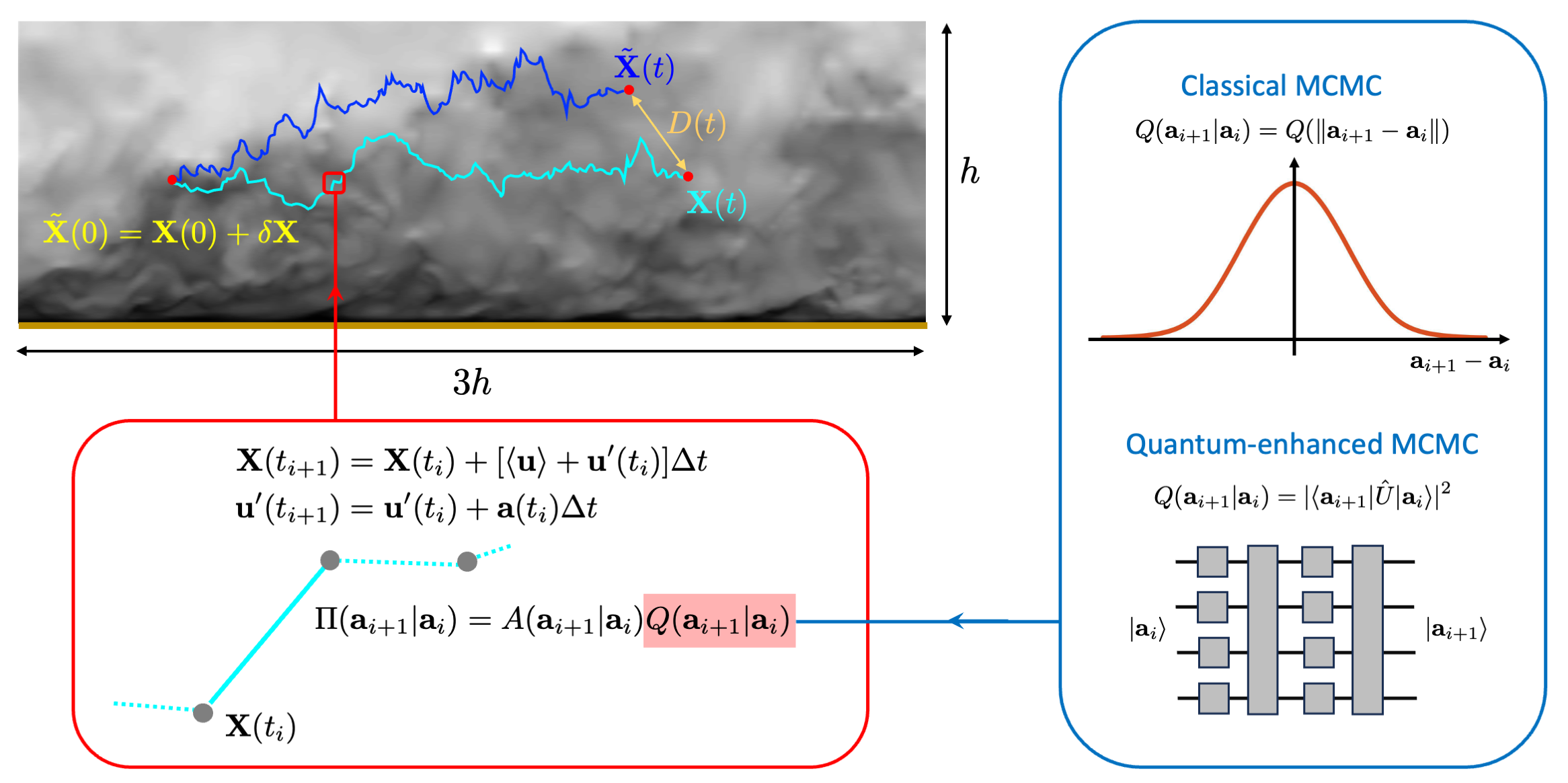}
    \caption{Principal sketch of the workflow of the Quantum-enhanced MCMC method for the synthetic generation of Lagrangian tracer trajectories in a turbulent boundary layer above the ground and its comparison with a classical MCMC method. The velocity fluctuations $\mathbf{u}'(\mathbf{x},t)$ about a mean turbulent shear flow $\langle \mathbf{u}\rangle$ are obtained from simulations. They are used to construct the target distribution of the acceleration $p(\mathbf{a})$ for the sampling. The contour plot (upper left) shows the turbulent kinetic energy of the velocity field, taken from a snapshot of the lower half of a channel simulation \cite{Graham2016}. Two initially very close tracer tracks are sketched. Their distance grows with time due to turbulent mixing. The red box below illustrates the stochastic particle advection method. The proposal distribution is either obtained classically or by the quantum algorithm, see blue box to the right. Note that the QE-MCMC algorithm is a hybrid quantum-classical algorithm.}
    \label{fig:principle}
\end{figure*}

Quantum computing methods in general \cite{Nielsen2010} are known for the efficient sampling properties, caused by the nature of the measurement process \cite{Arute2019,Wild2021}. Recently, the MCMC method was extended to a quantum-enhanced MCMC (QE-MCMC) algorithm by Layden et al. \cite{Layden2023}. The authors could show that by taking samples of the proposal distribution $Q(\mathbf{a}^*|\mathbf{a})$, which is constructed by a quantum algorithm, a speedup for a notoriously hard spin-glass Ising model problem is obtained. The target distribution $p(\mathbf{a})$ to sample from is a Boltzmann distribution in this case,
\begin{equation}
    p(\mathbf{a})= \frac{1}{\cal Z} \exp[-\hat H(\mathbf{a})/k_BT]\,,
\label{eq:canonical}    
\end{equation}
with the partition sum ${\cal Z}$ (which sums over all possible spin configurations $\mathbf{a}$), the Hamilton operator $\hat H(\mathbf{a})$ quantifying the energy stored in each specific spin-glass configuration, the Boltzmann constant $k_B$, and the temperature $T$. They showed that QE-MCMC gave the big\-gest spectral gap $\delta=1-|\lambda_2|$ between the first, $\lambda_1=1$, and second eigenvalue, $\lambda_2<\lambda_1$, of the Markov transition matrix $\Pi(\mathbf{a}^*|\mathbf{a})$ when compared to classical sampling, see Sec. II for further details. The width of the spectral gap is the essential quantity as it determines the convergence rate to the target distribution $p(\mathbf{a})$ of the Markov chain \cite{Levin2017}. The wider the gap the faster the convergence. In \cite{Layden2023}, a quadratic speed-up was obtained for the quantum physical problem compared to classical MCMC sampling. 

A quantum advantage is, however, not always guaranteed. In QE-MCMC, the assumed symmetry of the proposal distribution $Q$ is implemented in the unitary quantum circuit $\hat U$ with $\hat{U}^{-1}=\hat{U}^{\dagger}$ requiring in addition $\hat U=\hat U^\mathrm{T}$. This typically results in deep quantum circuits, which complicate a detailed understanding and tuning of the spectral gap $\delta$, as shown in ref. \cite{Nakano2024}. In addition, Orfi and Sels \cite{Orfi2024} constructed a sampling problem that does not obtain a quantum advantage over the classical counterpart. As a consequence, Nakano et al. \cite{Nakano2026} adapted the QE-MCMC algorithm by combination with a classical generative neural sampler (GNS) in the form of an autoencoder to accelerate the evaluation of the proposal distribution. In this hybrid algorithm, the calculation of $A(\mathbf{a}^*|\mathbf{a})$ was drastically simplified. In \cite{Ferguson2025}, a coarse grained QE-MCMC was presented to perform sampling tasks with considerably smaller qubit numbers. 

With a view to these open points and solution strategies, we want to investigate in the following the applicability of the original QE-MCMC algorithm to a sampling problem, which arises in classical fluid turbulence as a subtask next to the time advancement of the fluid flow equations. To this end, we apply the hybrid quantum-classical QE-MCMC method to draw acceleration field samples for modeling the Lagrangian material transport and dispersion of substances, described as a cloud of individual tracer particles in a turbulent fluid flow.  

We consider two application cases: (1) a homogeneous shear flow (HSF), which will serve as the benchmark case for the approach. This turbulent flow is characterized by a uniform mean shear rate $S = {\rm const.}$ \cite{Ingelmann2025}. (2) A turbulent pressure-driven shear flow, which is formed in a channel flow geometry \cite{Kim1987}. This flow develops turbulent boundary layers (TBL) and is characterized by a mean shear rate $S(y)$ that depends on the vertical distance $y$ from the horizontal channel wall. The substance to be transported is then considered as an ensemble of small and practically massless tracer particles, which nearly perfectly follow the turbulent fluid flow --- the Lagrangian perspective on fluid dynamics \cite{Yeung2002,Toschi2009,Sreenivasan2010,Mazzitelli2014,Mazzitelli2014a}. 

The QE-MCMC algorithm will generate synthetic particle tracks by sampling from a target distribution $p(a_x,a_y,a_z|\,y)$ of the three components of the acceleration vector field ${\bf a}$. Acceleration component statistics is inferred from the covariance of velocity fluctuations ${\bf u}'$, coupled in a two-layer stochastic model that contains two internal flow time scales \cite{Sawford1991,Viggiano2020}. A schematic illustration is shown in Fig.~\ref{fig:principle}. Each component of the acceleration vector field varies between a minimal and maximal value; the distributions also change with increasing distance from the wall $y$. This sets our multivariable sampling task. Our studies are compared with a classical MCMC algorithm and a stochastic tracer transport model. The quantum algorithm reproduces the particle pair dispersion $D(t)$ in agreement with the other classical methods; $D(t)$ is a central measure that quantifies the strength of turbulent mixing in the flow in a scaling law with respect to time. We also show that QE-MCMC method obeys a spectral gap $\delta$ that is comparable to the classical MCMC for qubit number per dimension $N_q=6$ in the HSF case; the (effective) spectral gap is significantly larger in the more complex TBL case for the quantum algorithm which supports the applicability of QE-MCMC for more complex sampling tasks. 

Most transport models for the dispersion of substances in the atmosphere are based on large-eddy simulations and discuss the scalar mixing in the Eulerian frame of reference \cite{Nieuwstadt1992,Dosio2006}. Classical stochastic transport models, as those discussed here, have been used to investigate the transport of chemical tracers or aerosols in the atmosphere in global circulation models \cite{Stein2015,Hoffmann2019}. Next to the resolved large-scale wind fields, unresolved velocity fluctuations at meso- and small scales are described as Gaussian fields, which, however, mostly neglect the cross-correlations between different components \cite{Hoffmann2016}. Our hybrid quantum-classical model goes beyond this stage by including cross correlations between the velocity and acceleration components in the corresponding covariances together with a large-scale correlation length and a two internal flow time scales in the underlying two-layer stochastic model, motivated by concepts of refs. \cite{Sawford1991,Viggiano2020}. 

Our adapted QE-MCMC method provides a quantum algorithm for the complementary Lagrangian description of fluid dynamics, where the frame of reference is connected to individual fluid particles. This opens new physical perspectives for understanding transport processes in fluid turbulence, as discussed above. We note here that basically all previous applications of quantum computing for fluid mechanics are formulated in the standard Eulerian frame of reference  \cite{Lubasch2020,Todorova2020,Gaitan2020,Budinski2021,Pfeffer2022,Pfeffer2023,Bharadwaj2023,Jaksch2023,Meng2023,Ingelmann2024,Ahmed2024,Pool2024,Brearley2024,Meng2024,Koecher2025,Bharadwaj2025,Over2025,Pfeffer2025}, see also refs. \cite{Bharadwaj2020,Succi2023a,Tennie2025} for reviews.

The outline is as follows. In Sec. II, we summarize the MCMC and QE-MCMC methods for completeness. Section III reports the results, first for a homogeneous shear flow as a benchmark, then for the turbulent channel flow. Finally, the work ends with a conclusion and an outlook.

\section{Methods}

\subsection{Classical Markov chain Monte Carlo method}
MCMC sampling is among the most popular algorithms to take samples from a targeted stationary probability density function $p(\mathbf{a})$ without explicitly computing it \cite{Bishop2006}. To this end, the algorithm creates a Markov chain, which implies that the next sample $\mathbf{a}^*$ depends only on the previous sample $\mathbf{a}$. The chain is specified by a transition matrix $\Pi(\mathbf{a}^*|\mathbf a)$. The convergence of the chain to the stationary target distribution $p(\mathbf{a})$ is satisfied when the {\it detailed balance condition} holds,
\begin{equation}
    p(\mathbf{a}^*) \Pi(\mathbf{a}^*|\mathbf{a})=p(\mathbf{a}) \Pi(\mathbf{a}|\mathbf{a}^*)\,.
\end{equation}
Furthermore, it is required that the chain is irreducible and aperiodic, two properties that are satisfied in standard cases. As already discussed in the introduction, a common strategy in MCMC algorithms is to decompose the Markov step into 2 substeps, a proposal and acceptance step. Off-diagonal elements of the transition matrix $\Pi$ can be written as \eqref{eq:Pi}. One widely used MCMC technique is the Metropolis-Hastings algorithm \cite{Metropolis1953,Hastings1970}, which constructs a Markov chain. The method proceeds iteratively as follows. At each iteration $n$, a new candidate point $\mathbf{a}^*$ is proposed based on the current state $\mathbf{a}^{(n)}$ using  proposal distribution $Q(\cdot\,|\mathbf{a}^{(n)})$. In our case, we use a simple isotropic Gaussian centered at the current state:
\begin{align}
    \mathbf{a}^* \sim Q(\cdot\,|\mathbf{a}^{(n)})
\end{align}
where
\begin{align}
Q(\mathbf{v}|\mathbf{a}) = \frac{1}{(2\pi \sigma_\mathrm{prop}^2)^{d/2}} \exp\left( -\frac{1}{2\sigma_\mathrm{prop}^2} \| \mathbf{v} - \mathbf{a} \|^2 \right).
\label{ProposalD}
\end{align}
Here, $\sigma_\mathrm{prop}^2$ denotes the proposal variance, which controls the size of the random perturbation applied to the current state. The choice of $\sigma_\mathrm{prop}$ is critical for the efficiency and convergence behavior of the Markov chain:
\begin{enumerate}
    \item[(1)] If $\sigma_\mathrm{prop}$ is too small, proposed moves will be very close to the current state. This leads to high acceptance rates but poor exploration of the state space, resulting in slow mixing and high autocorrelation between successive samples. The chain may require a very large number of iterations to approximate the target distribution well. 
    \item[(2)] If $\sigma_\mathrm{prop}$ is too large, proposed moves are likely to fall into low-probability regions of $p(\mathbf{a})$, and thus be rejected frequently. This leads to a low acceptance rate, causing the chain to remain stuck at its current position for many iterations, again impairing convergence.
\end{enumerate}
Once a proposal $\mathbf{a}^*$ has been generated, the next step is to decide whether to accept the proposed move. For this, we compute the acceptance probability:
\begin{align}
    A(\mathbf{a}^*|\mathbf{a}^{(n)}) &= \min\left\{1,\,\frac{p(\mathbf{a}^*)}{p(\mathbf{a}^{(n)})} \frac{Q(\mathbf{a}^{(n)}|\mathbf{a}^*)}{Q(\mathbf{a}^*|\mathbf{a}^{(n)})} \right\}\nonumber\\ 
    &= \min\left\{1,\,\frac{p(\mathbf{a}^*)}{p(\mathbf{a}^{(n)})} \right\},
\label{eq:class_symm}
\end{align}
where the simplification follows from the symmetry of the proposal distribution, $Q(\mathbf{v}|\mathbf{a}) = Q(\mathbf{a}|\mathbf{v})$, see again Eq. \eqref{ProposalD}.

Then, a uniform random number ${\cal U} \sim \operatorname{Uniform}(0,1)$ is drawn. The proposed sample is accepted if ${\cal U} \leq A(\mathbf{a}^* \mid \mathbf{a}^{(n)})$, and rejected otherwise:
\begin{align}
    \mathbf{a}^{(n+1)} = \begin{cases} 
        \mathbf{a}^* & \text{if\; } {\cal U} \leq A(\mathbf{a}^* \mid \mathbf{a}^{(n)}) \\ \mathbf{a}^{(n)}, & \text{otherwise}.
    \end{cases}
\end{align}
This accept–reject rule preserves the detailed balance condition, ensuring that the Markov chain converges to the correct equilibrium distribution, and therefore enables accurate sampling from the target distribution $p(\mathbf{a})$.

The performance of the classical MCMC is evaluated by the rate of convergence to $p(\mathbf{a})$. It can be obtained by a so-called total variation distance (TVD) $\xi$. The stationary distribution $p(\mathbf{a})$ is linked with an eigenvalue of $\lambda_1=1$ of $\Pi$. The spectrum of $\Pi$ is given by $\lambda_1>\lambda_2> \lambda_3 > ...$ Essential for the convergence rate is the spectral gap between $\lambda_1$ and $\lambda_2$ which is denoted as $\delta=1-|\lambda_2|$. In a nutshell, this gap quantifies the deviations to the stationary distribution. TVD and spectral gap can be shown to bound the mixing time $t_{\rm mix}$ as follows \cite{Levin2017}:
\begin{equation}
    \left(\frac{1}{\delta}-1\right)\,\log\left(\frac{1}{2\xi}\right) \le t_{\rm mix} \le \frac{1}{\delta}\,\log\left(\frac{1}{2\xi\, p_{\rm min}}\right)\,,
\end{equation}
with $p_{\rm min}=\min_{\mathbf{a}} p(\mathbf{a})$. In other words, the bigger the gap the faster the MCMC method is converged and the shorter the mixing time $t_{\rm mix}$ becomes.

\begin{figure}[t]
    \centering
    \includegraphics[width=0.49\textwidth]{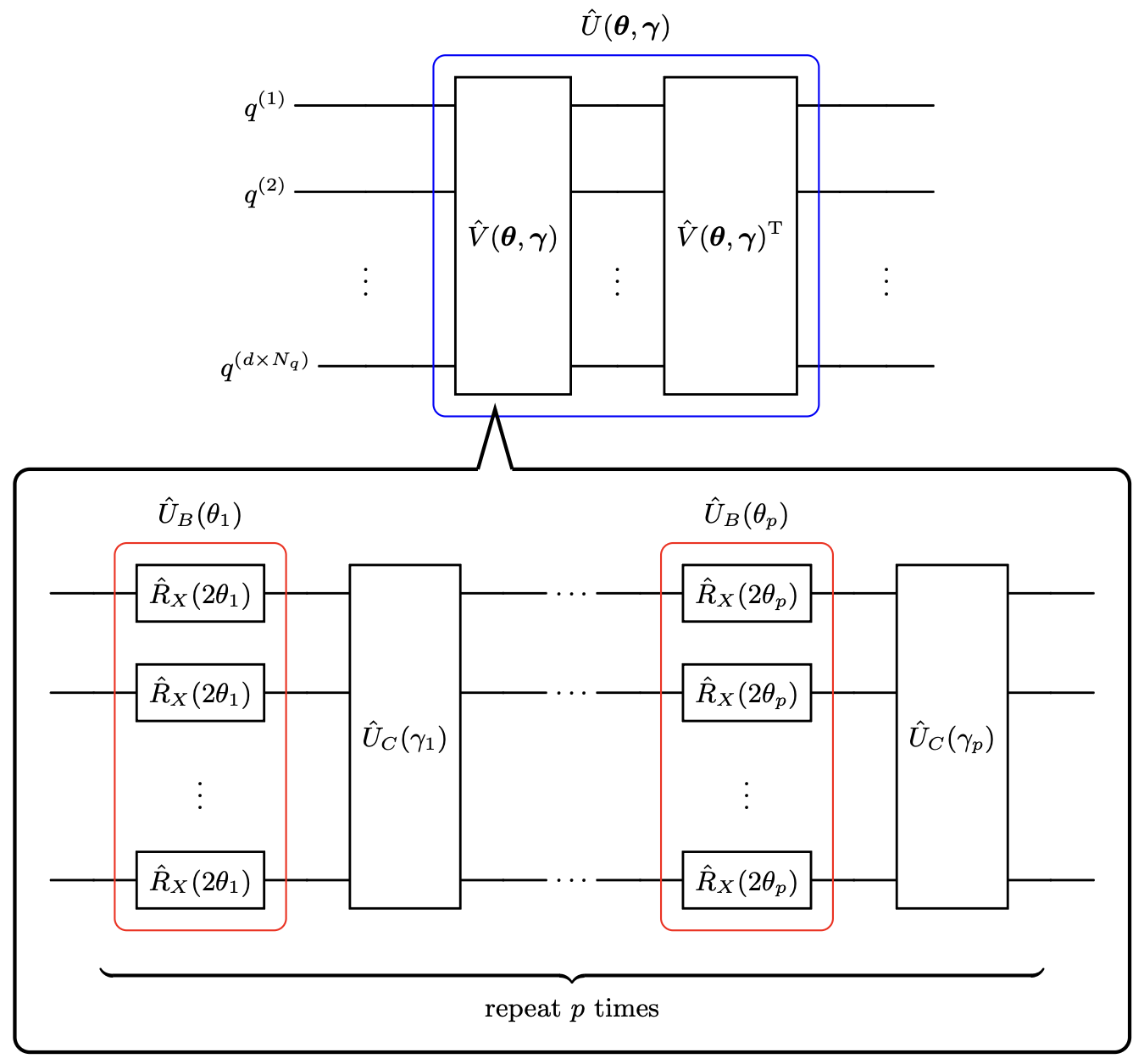}
    \caption{Quantum circuit of the distribution $Q$ sketched with $d\times N_q$ qubits. The operator $\hat V$ consists of $p$ layers of rotation operators $\hat R_X(2\theta)$ followed by a unitary $\hat U_C(\gamma)$ that entangles all qubits in each layer and follows from $\hat{H}_{\rm prob}$. The layers of $\hat R_X$ are connected to $\hat{H}_{\rm mix}$.}
    \label{fig:quantum_circuit}
\end{figure}

\subsection{Quantum-enhanced Markov chain Monte Carlo method}
The quantum-enhanced MCMC method (QE-MCMC) is a hybrid quantum-classical algorithm that operates on the classical data. The new sample $\mathbf{a}^*$ is suggested by a parametric quantum circuit in a unitary evolution step, i.e.,
\begin{equation}
    Q(\mathbf{a}^*|\mathbf{a})=|\langle \mathbf{a}^{*}|\hat U|\mathbf{a}\rangle|^2\,,
\label{eq:q_quantum}    
\end{equation}
with $\hat{U}^{-1}=\hat{U}^{\dagger}$. In short, the proposal step is quantum. For the quantum case, we will use the bra-ket notation for the statevectors \cite{Nielsen2010}. Typically, one takes $\hat U=\exp(-i\hat Ht)$ with $\hat H=\hat H^{\dagger}$. Operator $\hat H$ is the Hamiltonian of the quantum system that is supposed to take the quantum task. 

QE-MCMC is a class of quantum-inspired sampling algorithms that combine ideas from quantum computing with classical Metropolis-Hastings steps to explore complex probability landscapes more efficiently. As already said, the central idea is to replace the classical proposal distribution with one derived from a quantum circuit, whose measurement outcomes define the sampling candidates. Importantly, while the proposal distribution is now quantum-inspired, the overall structure of the Metropolis-Hastings algorithm—including the accept–reject rule—remains unchanged.

At each iteration $n$, a new $\mathbf{a}^*$ is proposed according to a transition probability
\begin{align}
    \mathbf{a}^* \sim Q(\cdot\,|\mathbf{a}^{(n)}), 
\end{align}
where $Q(\mathbf{a}^*|\mathbf{a}^{(n)})$ is given by \eqref{eq:q_quantum}. The transition probability corresponds to the probability that a quantum system, initialized in $|\mathbf{a}^{(n)}\rangle$, collapses into state $|\mathbf{a}^*\rangle$ after application of the unitary $\hat U$ and a measurement. The quantum circuit itself consists of alternating unitary transformations associated with two Hamiltonians that are combined with each other: a mixing Hamiltonian $\hat H_\mathrm{mix}$ and a cost (or problem) Hamiltonian $\hat H_\mathrm{prob}$ \cite{Layden2023,Nakano2024}. The general form of the circuit is given by
\begin{align}
    \hat U = \hat V(\boldsymbol\theta, \boldsymbol\gamma)\,\hat V(\boldsymbol\theta, \boldsymbol\gamma)^\mathrm{T} 
\end{align}
with
\begin{align}\label{eq:V_proposal}
\hat V(\boldsymbol\theta, \boldsymbol\gamma) = \prod_{j=1}^{p} \hat U_C(\gamma_j) \, \hat U_B(\theta_j),
\end{align}
where 
\begin{align}
    \hat U_B(\theta_j) &= \exp(-i \theta_j \hat H_\mathrm{mix}), \\
    \hat U_C(\gamma_j) &= \exp(-i \alpha \gamma_j \hat H_\mathrm{prob}), \label{eq:U_C_hat}
\end{align}
and $p$ is a hyperparameter that determines the depth of the
circuit. The constant $\alpha = |\hat{H}_\mathrm{mix}|_F / |\hat{H}_\mathrm{prob}|_F$ in Eq.~\eqref{eq:U_C_hat} ensures that the contributions of the mixing and problem Hamiltonians enter at comparable scales, thereby balancing exploration and exploitation in the quantum proposal ($| \cdot |_F$ denotes the Frobenius norm). The layered structure in Eq.~\eqref{eq:V_proposal} is inspired by the Quantum Approximate Optimization Algorithm (QAOA) \cite{Nakano2024}, and the parameters $\boldsymbol\theta=\{\theta_1,\dots,\theta_p\}$ and $\boldsymbol\gamma=\{\gamma_1,\dots,\gamma_p\}$ control the interplay between exploration and exploitation in the quantum proposal. To reduce the complexity of the parameter space, we consider here a simplified circuit with a single effective parameter $\theta$, and define
\begin{align}
\hat U = \hat V(\theta) \, \hat V^\mathrm{T}(\theta) \;\;\; \text{with} \;\;\; \hat V(\theta) = [\hat U_C(\theta) \, \hat U_B(\theta)]^p.
\end{align}
By construction, this unitary $\hat U$ is symmetric, i.e., $\hat U = \hat{U}^\mathrm{T}$, ensuring a symmetric proposal distribution. This {\em additional} symmetry of the unitary guaranties cancelation of the proposal terms in the acceptance ratio, simplifying the algorithm. 
The quantum circuit used to construct the proposal is schematically illustrated in Fig. \ref{fig:quantum_circuit}. The mixing Hamiltonian is chosen as
\begin{align}
    \hat{H}_\mathrm{mix} = \sum_{j=1}^{d\times N_q} \hat X_j,
\end{align}
where $N_q$ denotes the number of qubits per dimension $d$, and $\hat X_j$ are Pauli $X$ operators acting on qubit $j$, inducing transitions between bit strings. The problem Hamiltonian encodes the effective energy landscape of the target distribution:
\begin{align}
    \hat{H}_\mathrm{prob} = \sum_{\mathbf{a}} E(\mathbf{a}) \, |\mathbf{a}\rangle \langle \mathbf{a}|,
\end{align}
where 
\begin{align}
E(\mathbf{a}) = \mathbf{a}^\mathrm{T} \, \hat\Sigma^{-1} \, \mathbf{a}.
\end{align}
Here, $E(\mathbf{a})$ should not be interpreted as a physical energy, but rather as a Mahalanobis-type squared acceleration distance that quantifies how atypical a given acceleration vector is with respect to the covariance $\hat\Sigma$. 

After preparing the quantum state via the circuit $\hat U$, a measurement in the computational basis is performed. This measurement collapses the quantum state into a classical bit string $\mathbf{a}^*$, sampled with probability $Q(\mathbf{a}^*|\mathbf{a}^{(n)})$. Since quantum measurement always yields one of the $2^{d\times N_q}$ basis states, the output samples $\mathbf{a}^*$ are discrete vectors in $\{0,1\}^{d\times N_q}$. Notably, this is a {\em one-shot quantum algorithm} which avoids repeated runs to reconstruct all probabilities of the components of the quantum state. Thus, each qubit contributes one bit of resolution in the sample space. To apply this scheme to continuous target distributions, the state vector $\mathbf{a}$ must first be discretized and assigned to a bit string, and the corresponding inverse mapping must be applied after sampling.

The candidate $\mathbf{a}^*$ is then accepted or rejected according to a Metropolis-like rule. A uniform random number ${\cal U} \sim \operatorname{Uniform}(0,1)$ is drawn, and the acceptance probability is computed as:
\begin{align}
A(\mathbf{a}^*|\mathbf{a}^{(n)}) = \min\left\{1,\,\frac{p(\mathbf{a}^*)}{p(\mathbf{a}^{(n)})} \, \frac{Q(\mathbf{a}^{(n)}|\mathbf{a}^*)}{Q(\mathbf{a}^*|\mathbf{a}^{(n)})} \right\}.
\end{align}
Due to the additional symmetry $\hat{U}=\hat{U}^\mathrm{T}$ of the quantum circuit, 
\begin{align}
Q(\mathbf{a}^*|\mathbf{a}^{(n)})&=
|\langle\mathbf{a}^*|\hat U|\mathbf{a}^{(n)}\rangle|^2\nonumber\\
&=|\langle\mathbf{a}^{(n)}|\hat U|\mathbf{a}^{*}\rangle|^2
= Q(\mathbf{a}^{(n)}|\mathbf{a}^*),
\end{align}
the acceptance rule simplifies to a Metropolis step, cf. Eq.\eqref{eq:class_symm}. The rule is then given by
\begin{align}
    A(\mathbf{a}^*|\mathbf{a}^{(n)}) = \min\left\{1,\,\frac{p(\mathbf{a}^*)}{p(\mathbf{a}^{(n)})} \right\}.
\end{align}

In our setting, the target distribution is a multivariate normal distribution, $p(\mathbf{a}) \propto \exp(-\tfrac{1}{2} \mathbf{a}^\mathrm{T} \hat\Xi^{-1} \mathbf{a})$. This can be written in Boltzmann form as $p(\mathbf{a}) \propto \exp(-E(\mathbf{a})/(k_B T))$. The Boltzmann constant $k_B=1$ in the following. The effective energy becomes $E(\mathbf{a}) = \mathbf{a}^\mathrm{T} \hat\Sigma^{-1} \mathbf{a}$, provided that $\hat\Xi = \tfrac{T}{2}\hat\Sigma$. In the reference case corresponding to the original Gaussian, this yields $T=2$. 

For additional flexibility, we consider sampling temperatures $T>2$ in QE-MCMC and treat $T$ as a tunable parameter. This corresponds to sampling from a tempered target distribution, $p_T(\mathbf{a}) \propto \exp(-E(\mathbf{a})/T)$, which broadens the distribution and increases the probability of visiting less likely states. As a consequence, the QE-MCMC chain no longer samples exactly the original target distribution, but instead a controlled, temperature-modified variant. This allows us to adjust the balance between exploration and exploitation, as discussed further in Sec.~III.

This leads to the final expression for the acceptance probability
\begin{align}
A(\mathbf{a}^*|\mathbf{a}^{(n)}) = \min\left\{1,\,\exp\left( -\frac{E(\mathbf{a}^*) - E(\mathbf{a}^{(n)})}{T} \right) \right\}\,.
\label{eq:acceptance}
\end{align}
The Markov chain is updated accordingly:
\begin{align}
\mathbf{a}^{(n+1)} =
\begin{cases}
\mathbf{a}^*, & \text{if\;\;} {\cal U} \leq A(\mathbf{a}^*|\mathbf{a}^{(n)}) \\
\mathbf{a}^{(n)}, & \text{otherwise}.
\end{cases}
\end{align}
This hybrid algorithm retains the convergence guarantee of classical MCMC while exploiting quantum coherence and interference for nonlocal and potentially more efficient exploration of the state space.

\section{Results}
In the following, the application of the QE-MCMC algorithm to two turbulent fluid flow configurations is presented. We will start with the simplest extension to homogeneous isotropic turbulence in a cubic box, the homogeneous shear flow case \cite{Shen1997,Pope2002,Schumacher2001,Schumacher2004}. This configuration has already been investigated by Ingelmann et al. \cite{Ingelmann2025}. Here, it serves as a benchmark for the new QE-MCMC application. The complexity of the application case increases subsequently by turning to a turbulent channel flow with near-wall turbulent boundary layers \cite{Pope2000,Graham2016,Polanco2018}. Here, we used the simulation data record for a turbulent channel flow at a friction Reynolds number $Re_{\tau}=1000$ \cite{Graham2016}, which is provided in the Johns Hopkins Turbulence Database \cite{Li2008}.   

\subsection{Homogeneous shear flow}

\subsubsection{Flow configuration}
MCMC and QE-MCMC methods were first tested in a three-dimensional homogeneous turbulent shear flow, see also ref. \cite{Ingelmann2025}. Such a turbulent flow, which evolves far away from walls and boundaries, is commonly described by a decomposition of the velocity field into a streamwise mean flow and fluctuations which are given by
\begin{align}
\mathbf{u}(\mathbf{x}, t) = \langle u_x\rangle(y) \mathbf{e}_x + \mathbf{u}^\prime(\mathbf{x}, t)\,. 
\end{align}
The homogeneous mean flow is given by
\begin{align}
\langle u_x\rangle(y) = Sy\,,
\end{align}
where $\langle u_x\rangle$ denotes the mean streamwise velocity, linearly increasing with the wall-normal coordinate $y$ with a constant shear rate $S$. Note that the indices $i,j,k$ take values of $\{x,y,z\}$, which correspond to $\{1,2,3\}$. The vector $\mathbf{u}^\prime$ contains the velocity fluctuations around this mean profile. Due to the imposed shear, correlations arise between the fluctuating velocity components $u_x^\prime$ and $u_y^\prime$, which are particularly important for the evolution of passive tracers or particles embedded in the flow.

The motion of such particles is modeled using a system of stochastic differential equations (SDE) of Langevin type. Differently to ref. \cite{Ingelmann2025}, we extend the stochastic model in this work by two points to model a more realistic particle dispersion process:  
\begin{enumerate}
\item[(1)] Streamwise extended velocity structures with a finite length develop in both types of shear flows, streamwise streaks in the homogeneous shear flow \cite{Schumacher2000,Gualtieri2002} as well as  large-scale motions or turbulent superstructures in high-Reynolds-number turbulent channel flows \cite{Bailey2010,Smits2013}. Thus, we include an exponential correlation decay term with a characteristic length scale $\ell_{\rm corr}$ in the large-scale component of the flow, similar to ref. \cite{Mazzitelli2014a}.
\item[(2)] Turbulence is characterized by multiple time scales. Thus, we extend the basic stochastic model to a two-equation model for velocity fluctuations and accelerations, which was originally introduced by Sawford \cite{Sawford1991} and recently refined and extended by Viggiano et al. \cite{Viggiano2020}. 
\end{enumerate}
Both steps in combination will lead to more realistic tracer dispersion behavior in the Lagrangian turbulence. The equations for each tracer particle pair $(k)$ (which should not be mixed with the index of a Markov chain step) are given in variables $\{X_i, u'_i,a_i\}$ for tracer 1 and $\{\tilde{X}_i, \tilde{u}'_i,\tilde{a}_i\}$ for tracer 2. Here, we write down those of tracer 1 of pair $(k)$ only,  
\begin{align}
\mathrm{d}a_i^{(k)} &= -\frac{a_i^{(k)}}{\tau_{\eta}}\, \mathrm{d}t + G_{ij} \mathrm{d}W^{(k)}_j, 
\label{SDE1a}\\
\mathrm{d}u_i^{\prime\,(k)} &= -\frac{u_i^{\prime\,(k)}}{\tau_i}\, \mathrm{d}t + a_i^{(k)} \mathrm{d}t , 
\label{SDE1b}\\
\mathrm{d}X^{(k)}_i &= \left[ S Y^{(k)} {\cal F}(X^{(k)}, \ell_{\rm corr}) \delta_{ix} + u_i^{\prime (k)} \right] \mathrm{d}t,
\label{SDE1c}
\end{align}
with the spatial correlation decay term
\begin{align}
{\cal F}(X^{(k)}, \ell_{\rm corr})= \exp\left(-\frac{|X^{(k)}-\tilde{X}^{(k)}|}{\ell_{\rm corr}}\right)
\label{SDE1d}
\end{align}
where $\tau_i$ denotes a characteristic relaxation time scale, $\tau_{\eta}<\tau_i$ the Kolmogorov time scale, $G_{ij}$ a diffusion tensor; $\mathrm{d}W_j$ represents independent Wiener processes and $\ell_{\rm corr}$ is a characteristic spatial correlation length scale controlling the exponential decay of $\cal F$ with the streamwise tracer separation \cite{Sawford1991,Pope2002}. The first pair of equations governs the random acceleration components $a_i(t)$ and $\tilde{a}_i(t)$ of both tracers in the pair, the second their evolution of the fluctuating velocity components $u_i^\prime(t)$ and $\tilde{u}_i'(t)$, and the third describes the corresponding particle positions $X_i(t)$ and $\tilde{X}_i(t)$. The term $S Y {\cal F}(X^{(k)},\ell_{\rm corr}) \delta_{ix}$ reflects the deterministic advection due to the mean shear profile in the streamwise direction $x$. Note that upper case letters $(X,Y,Z)=(X_1,X_2,X_3)$ are used for spatial coordinates $X_i$ of the Lagrangian tracer particles in Eqs. \eqref{SDE1c} and \eqref{SDE1d} (not mixed with Pauli operator $X$); the same holds for tracer 2 in the pair. In discrete time $t_n=n\Delta t$ with a fixed $\Delta t$, the system \eqref{SDE1a}--\eqref{SDE1c} becomes
\begin{align*}
   a_i^{(k)}(t_{n+1}) = & a_i^{(k)}(t_n) - \frac{a^{(k)}_i(t_n)}{\tau_\eta}\,\Delta t + G_{ij}\,\Delta W^{(k)}_j(t_n), \\
   u_i^{\prime (k)}(t_{n+1}) = & u_i^{\prime (k)}(t_n) - \left[\frac{u_i^{\prime (k)}(t_n)}{\tau_i}-a_i^{(k)}(t_n)\right]\Delta t, \\                 
   X_i^{(k)}(t_{n+1}) =& X_i^{(k)}(t_n) \, + \\
                                     &\left[ S Y^{(k)}(t_n) {\cal F}(t_n) \delta_{ix} + u_i^{\prime (k)}(t_n) \right] \Delta t.
\end{align*}
Numerically, it is solved by an Euler-Murayama scheme \cite{Higham2021}. We use the Einstein summation convention. 

To close the stochastic model, the second-order velocity statistics are prescribed through a Gaussian description of the velocity fluctuations $\mathbf{u}'$. In the present setting, we assume a multivariate normal distribution with symmetric covariance matrix $\hat\Sigma_{\mathbf{u}'\mathbf{u}'}$, inferred from the shear-flow statistics of the simulation data,
\begin{equation}
    \hat\Sigma_{\mathbf{u}'\mathbf{u}'}
    =
    \begin{bmatrix}
    \Sigma_{xx} & \Sigma_{xy} & 0 \\
    \Sigma_{xy} & \Sigma_{yy} & 0 \\
    0 & 0 & \Sigma_{zz}
    \end{bmatrix}.
\end{equation}
This yields the bivariate in-plane density
\begin{equation}
    p^{(\mathbf{u}^\prime)}_{\rm 2D}(u_x',u_y')
    =
    \frac{1}{2\pi\,\bigl|\hat\Sigma^{(\mathbf{u})}_{\rm 2D}\bigr|^{1/2}}
    \exp\!\left(
    -\frac{1}{2}\,\mathbf{u}'^{\,\mathrm{T}}\,\bigl(\hat\Sigma^{(\mathbf{u}')}_{\rm 2D}\bigr)^{-1}\mathbf{u}'
    \right),
    \label{eq:Pxy_HSF_u}
\end{equation}
with
\begin{equation}
    \mathbf{u}'=
    \begin{bmatrix}
    u_x'\\ u_y'
    \end{bmatrix},\qquad
    \hat\Sigma^{(\mathbf{u}^\prime)}_{\rm 2D}=
    \begin{bmatrix}
    \Sigma_{xx} & \Sigma_{xy}\\
    \Sigma_{xy} & \Sigma_{yy}
    \end{bmatrix},
\end{equation}
and the statistically independent spanwise component is modeled by the univariate normal density
    \begin{equation}
    p^{(\mathbf{u}^\prime)}_{\rm 1D}(u_z')
    =
    \frac{1}{\sqrt{2\pi\,\Sigma_{zz}}}
    \exp\!\left(-\frac{u_z'^2}{2\Sigma_{zz}}\right).
    \label{Pz_HSF_u}
\end{equation}
While Eqs.~\eqref{SDE1a}--\eqref{SDE1c} evolve both $\mathbf{u}'$ and $\mathbf{a}$ dynamically, the MCMC and QE-MCMC variants require an explicit target distribution for the sampled variables. In this work, we sample the accelerations $\mathbf{a}$ at each time step. Direct acceleration statistics are not available from the data, hence we reconstruct the covariance $\hat\Sigma_{\mathbf{a}\mathbf{a}}$ from the diffusion tensor $\hat G$ in Eq.~\eqref{SDE1a}. In continuous time, the Ornstein-Uhlenbeck form \eqref{SDE1a} implies the stationary Lyapunov relation
\begin{align}
    \hat\Sigma_{\mathbf{a}\mathbf{a}} = \frac{\tau_\eta}{2}\,\hat G \hat G^\mathrm{T},
\label{eq:Sigmaaa_from_G}
\end{align}
which we use to define the acceleration covariance. The construction of $\hat G$ from $\hat\Sigma_{\mathbf{u}'\mathbf{u}'}$ is detailed in the appendix.

Analogously to the velocity model above, we assume a Gaussian factorization for the acceleration components in this framework and introduce the in-plane and spanwise acceleration densities
\begin{align}
    p^{(\mathbf{a})}_{\rm 2D}(a_x,a_y)
    &=
    \frac{1}{2\pi\,\bigl|\hat\Sigma^{(\mathbf{a})}_{\rm 2D}\bigr|^{1/2}}
    \exp\!\left(
    -\frac{1}{2}\,\mathbf{a}_{12}^\mathrm{T}\,\bigl(\hat\Sigma^{(\mathbf{a})}_{\rm 2D}\bigr)^{-1}\mathbf{a}_{12}
    \right),
    \\
    p^{(\mathbf{a})}_{\rm 1D}(a_z)
    &=
    \frac{1}{\sqrt{2\pi\,\Sigma^{(\mathbf{a})}_{zz}}}
    \exp\!\left(-\frac{a_z^{2}}{2\,\Sigma^{(\mathbf{a})}_{zz}}\right),
\end{align}
where
\begin{align}
    \mathbf{a}_{12}=
    \begin{bmatrix}
    a_x\\ a_y
    \end{bmatrix},\quad
    \hat\Sigma^{(\mathbf{a})}_{\rm 2D}=
    \begin{bmatrix}
    \Sigma^{(\mathbf{a})}_{xx} & \Sigma^{(\mathbf{a})}_{xy}\\
    \Sigma^{(\mathbf{a})}_{xy} & \Sigma^{(\mathbf{a})}_{yy}
    \end{bmatrix},\quad
    \Sigma^{(\mathbf{a})}_{ij}=(\hat\Sigma_{\mathbf{a}\mathbf{a}})_{ij}.
\end{align}
Accordingly, the target law used for acceleration sampling becomes
\begin{align}
    \mathbf{a}(t_n)\sim
    p^{(\mathbf{a})}_{\rm 2D}(a_x,a_y)\,\otimes\,p^{(\mathbf{a})}_{\rm 1D}(a_z).
    \label{tensorHSF}
\end{align}
The sampled accelerations are then inserted into the discrete-time update of $\mathbf{u}'$ and $\mathbf{X}$ (see Eqs.~\eqref{SDE1a}--\eqref{SDE1d}), providing a consistent two-stage model that couples the prescribed second-order velocity statistics $\hat\Sigma_{\mathbf{u}'\mathbf{u}'}$ to an inferred acceleration covariance $\hat\Sigma_{\mathbf{a}\mathbf{a}}$. Although real acceleration statistics are known to be non-Gaussian \cite{Toschi2009}, the Gaussian approximation in Eq.~\eqref{tensorHSF} enables a tractable construction of the two-equation model and a well-defined target distribution for MCMC and QE-MCMC sampling.

We note here that Eq.~\eqref{SDE1a} leads to an exponential decay of the temporal correlations of the acceleration,
\begin{equation}
C_i(\tau)=\langle a_i(t)a_i(t+\tau)\rangle \sim \exp(-\tau/\tau_\eta)\,.
\label{acc_corr}
\end{equation}
The Kolmogorov time $\tau_\eta$ determines the decay of the correlations. This is important for the present problem since the Markov chain is not an equilibrium sampler. This point will be detailed further in Sec.~III.A.3 when the optimal parameters of the MCMC methods are discussed.

\subsubsection{Lagrangian particle pair dispersion}\label{HSF_dispersion}
A key quantity of interest in the Lagrangian framework is the particle dispersion, which characterizes the spreading of particle trajectories due to turbulent mixing. To assess this quantity, we compute the mean-squared relative displacement $D(t)$ between pairs of Lagrangian particles, initialized with small differences in their initial conditions. Specifically, we define the dispersion in the $i$-th direction at time $t_n$ for all particle pairs $k$ as:
\begin{align}
D_i(t_n) &= \left\langle \left[\Delta X_i^{(k)}(t_n) - \Delta X_i^{(k)}(0)\right]^2 \right\rangle_L \notag \\
&= \frac{1}{N_\mathrm{pa}} \sum_{k=1}^{N_\mathrm{pa}} \left[ \Delta X_i^{(k)}(t_n) - \Delta X_i^{(k)}(0) \right]^2 
\label{LagDisp}
\end{align}
with
\begin{align}
\Delta X_i^{(k)}(t_n) = X_i^{(k)}(t_n) - \tilde{X}_i^{(k)}(t_n)\,,
\label{LagDisp1}
\end{align}
where $N_\mathrm{pa}$ denotes the number of sampled particle pair trajectories, and $X_i^{(k)}(t_n),\,\tilde X_i^{(k)}(t_n)$ are the trajectories of two particles from the $k$-th pair realization. The tracer pairs are initialized with a small displacement in the wall-normal direction $y$ as well as distinct initial accelerations $\mathbf{a}^{(0)}$ and velocity fluctuations $\mathbf{u}^{\prime\,(0)}$. Specifically, the joint initial state is sampled as
\begin{equation}
    \bigl(\mathbf{a}^{(0)},\,\mathbf{u}^{\prime\,(0)}\bigr)^{\mathrm T}\sim \mathcal{N}\!\left(\mathbf{0},\,\hat\Sigma_{\mathbf{z}\mathbf{z}}\right),
\end{equation}
with the block covariance matrix
\begin{equation}
    \hat\Sigma_{\mathbf{z}\mathbf{z}}
    =
    \begin{pmatrix}
    \hat\Sigma_{\mathbf{u}'\mathbf{u}'} & \hat\Sigma_{\mathbf{u}'\mathbf{a}}\\
    \hat\Sigma_{\mathbf{a}\mathbf{u}'} & \hat\Sigma_{\mathbf{a}\mathbf{a}}
    \end{pmatrix}
    \in \mathbb{R}^{6\times 6}.
\end{equation}
The construction of the individual blocks of $\hat\Sigma_{\mathbf{z}\mathbf{z}}$ from the prescribed shear-flow statistics is described the in appendix.

\begin{figure*}[t]
    \centering
    \includegraphics[width=0.8\textwidth]{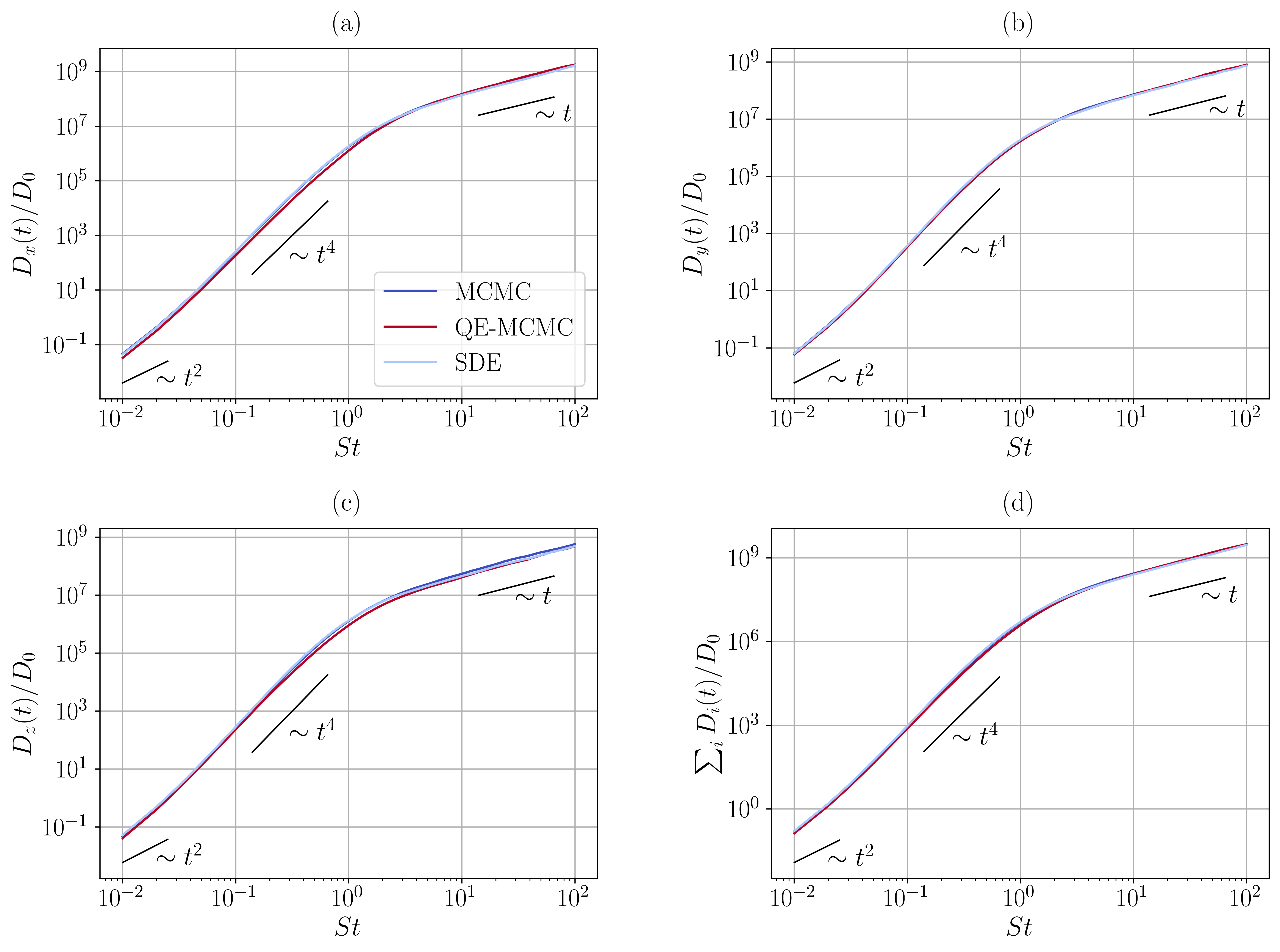}
    \caption{Comparison of dispersion curves computed using the solution of the stochastic differential equations \eqref{SDE1a}--\eqref{SDE1d} (SDE), the classical MCMC and QE-MCMC scheme. In all cases, the dispersion curves have been obtained for 3000 individual Lagrangian particle tracks, which were initialized with slightly different initial vertical positions. (a) Streamwise pair dispersion $D_x(t)$, (b) wall-normal pair dispersion $D_y(t)$, (c) spanwise pair dispersion $D_z(t)$, and (d) total pair dispersion $D(t)$.}
    \label{fig:dispersion_curves_shear_flow}
\end{figure*} 

We use the initial particle dispersion as a normalization constant. It is calculated according to
\begin{align}
    D_0 = \sum_{i=1}^3 \left\langle \left[\Delta X_i^{(k)}(0)\right]^2 \right\rangle_L = \frac{1}{N_\mathrm{pa}}\sum_{i=1}^3\sum_{k=1}^{N_\mathrm{pa}}\left[\Delta X_i^{(k)}\right]^2\,.
\end{align}

In the numerical benchmark experiments, the following simulation parameters were employed, equal to \cite{Ingelmann2025}, which follow from \cite{Pope2002} and the simulation data described therein. All quantities are given in characteristic time and length units. The shear rate was set to $S = 3.51$ inverse time units (and thus results in a shear time $S^{-1}=0.28$), and the time step used for time integration was $S\Delta t = 0.01$, and the mean kinetic energy dissipation rate $\varepsilon=1.34$. The Kolmogorov time $\tau_{\eta}=\sqrt{\nu/\varepsilon}\approx 0.07$ which is estimated from the Taylor microscale Reynolds number that varied between 40 and 110. As the decorrelation length $\ell_{\rm corr}=\sqrt{\varepsilon/S^3}\approx 0.18$ is taken in the present case. The number of time steps was set to $N_\mathrm{steps} = 10\,000$, resulting in a total simulation time of $ST=100$. The calculations were carried out for a number of $N_\mathrm{pa} = 3000$ particles. 

For the QE-MCMC simulation, we used $N_q=6$ qubits. The covariance matrix of the velocity fluctuations, the diffusion tensor and the relaxation times for the Langevin model were prescribed as detailed in \cite{Ingelmann2025}. The dispersion in each direction was computed using Eq. \eqref{LagDisp1}, and the total dispersion was obtained as the sum of the dispersion in the three spatial directions, $D(t) = \sum_{i=1}^3 D_i(t)$. The resulting dispersion curves are presented in Fig. \ref{fig:dispersion_curves_shear_flow}. For the smallest times $t\ll S^{-1}$, the particle dispersion comes out of a ballistic scaling regime with $D_i(t)\sim t^2$ for times $t\ll S^{-1}$. This is indicated here only, since the time step was taken larger to capture the long-term dispersion. Up to times $t\lesssim S^{-1}$, the scaling crosses over into a scaling with $D_i(t)\sim t^4$. For times $t\gtrsim S^{-1}$, spatial and temporal correlations are decayed such that a diffusive scaling is observed, $D_i(t)\sim t$. All three methods, both classical and the quantum-enhanced method, give practically the same results.

\subsubsection{Optimal parameters of the MCMC method}
To ensure efficient sampling within the MCMC framework, a parameter study was conducted to determine the optimal proposal variance $\sigma_\mathrm{prop}$ for the acceleration fluctuations, see again Eq.~\eqref{ProposalD} for $Q(\mathbf{a}^*|\mathbf{a}^{(n)})$. The proposal standard deviations are calibrated by matching the temporal correlation structure of the Markov chain to the target dynamics implied by the stochastic acceleration model. 
This calibration directly determines the mixing behaviour of the Markov chain, since the proposal scale controls the balance between exploration of the continuous state space and persistence of temporal correlations in the sampled trajectories.

In classical MCMC theory, the spectral gap of the Markov operator is often interpreted as an indicator of convergence speed, with larger values corresponding to faster relaxation towards the stationary distribution. However, this interpretation is only meaningful when the Markov chain is used as a purely equilibrium sampler. In the present case, the Markov chain is embedded into the temporal evolution of the stochastic acceleration process and therefore contributes directly to the generation of a time-ordered physical signal. As a consequence, its mixing properties cannot be optimized independently of the temporal correlation structure prescribed by the underlying model.

Since the acceleration process exhibits a characteristic relaxation time scale $\tau_\eta$, as shown in Eq.~\eqref{SDE1a}, the theoretical lag-$k$ autocorrelation is
\begin{equation}
    \rho_k^\mathrm{target}=\exp\!\left(-k\,\frac{\Delta t}{\tau_\eta}\right),\qquad k=1,\dots,K,
\end{equation}
where $\Delta t$, again, denotes the (physical) time step of the particle integrator and $K$ is the maximum lag considered. For a given candidate proposal scale $\sigma_\mathrm{prop}$, an MCMC chain is generated for the corresponding acceleration components, empirical autocorrelations $\rho_k(\sigma_\mathrm{prop})$ are computed, and the mismatch is quantified via a weighted least-squares error over the first $K$ lags,
\begin{equation}
    E(\sigma_\mathrm{prop})=\sum_{k=1}^{K} w_k\bigl(\rho_k(\sigma_\mathrm{prop})-\rho_k^\mathrm{target}\bigr)^2,
    \quad w_k=\frac{1}{k}.
\end{equation}
Because the coupled in-plane update $(a_1,a_2)$ or $(a_x,a_y)$ and the scalar update $a_3$ or $a_z$ generally exhibit different mixing behavior, two separate proposal scales are determined: $\sigma_{12,\mathrm{prop}}$ for the bivariate chain $(a_1,a_2)$ and $\sigma_{3,\mathrm{prop}}$ for the univariate chain $a_3$. For $(a_1,a_2)$, the objective combines the lagged autocorrelation errors of both components,
\begin{align}
    E_{12}(\sigma_{12,\mathrm{prop}})=
    \sum_{k=1}^{K} w_k\Bigl[&
    \bigl(\rho_{k}^{(a_1)}-\rho_k^\mathrm{target}\bigr)^2 \,+ \notag\\
    &\bigl(\rho_{k}^{(a_2)}-\rho_k^\mathrm{target}\bigr)^2 \Bigr],
\end{align}
while for $a_3$ an analogous scalar expression is used,
\begin{equation}
    E(\sigma_{3,\mathrm{prop}})=\sum_{k=1}^{K} w_k\bigl(\rho_k^{(a_3)}-\rho_k^\mathrm{target}\bigr)^2.
\end{equation}
To reduce statistical noise, each candidate proposal scale is evaluated by averaging the error over multiple independent repeats, and the optimal values are obtained by a grid search,
\begin{align}
    &\sigma_{12,\mathrm{prop}}^*=\arg\min_{\sigma_{12,\mathrm{prop}}} E_{12}(\sigma_{12,\mathrm{prop}}), \\
    &\sigma_{3,\mathrm{prop}}^*=\arg\min_{\sigma_{3,\mathrm{prop}}} E_{3}(\sigma_{3,\mathrm{prop}}).
\end{align}
For the parameter set considered here, we set $K=5$, which yielded $\sigma_{12,\mathrm{prop}}^*=1.75$ for the coupled $(a_1,a_2)$ update and $\sigma_{3,\mathrm{prop}}^*=2.00$ for the scalar $a_3$ update. At these values, the distribution $Q$ produces steps of appropriate magnitude, enabling efficient exploration of the state space while reproducing the intended temporal correlation of the target stochastic dynamics. Consequently, the induced particle-trajectory statistics $X_i(t)$ are consistent with the intended stochastic dynamics.

Finally, it is worth noting that the spectral gap of the resulting Markov chain remains a useful diagnostic quantity for characterizing mixing efficiency in the classical MCMC sense. In particular, it provides information about the relaxation properties of the chain in the abstract sense of convergence towards its stationary distribution. However, in the present setting it should not be interpreted as an unconditional figure of merit or as an independent optimization objective. This is because the target distribution is not tuned to maximize mixing speed alone, but is instead calibrated to reproduce the prescribed temporal correlation structure of the underlying stochastic acceleration model on the time scale $\tau_\eta$. As a consequence, the spectral gap reflects a constrained trade-off between two competing requirements: efficient exploration of the continuous state space, and preservation of the physically relevant temporal dependence structure of the generated trajectories. In this sense, the spectral gap provides a meaningful but conditional measure of performance, which must be interpreted in conjunction with the autocorrelation-based calibration criterion used to determine the proposal parameters.

\subsubsection{Optimal parameters of the QE-MCMC method}
The quantum-enhanced MCMC introduces additional parameters that influence the structure and efficiency of the sampling process. In the present setting, QE-MCMC was applied to the distribution of the acceleration fluctuations using separate quantum proposals for the joint ($a_1$, $a_2$) and individual spanwise ($a_3$) components. The corresponding distributions are not accessible from the DNS data records, thus they will be reconstructed from the Gaussian velocity fluctuation statistics. 

For each of these components, QE-MCMC employs a quantum walk-based proposal mechanism, which is governed by the following tunable parameters.
\begin{enumerate}
    \item [(1)] {\em Temperature parameters $T_{12}$ and $T_3$} control the exploration scale of the tempered target distribution. Higher values result in broader range of proposal values, thus enhancing exploration, but potentially reducing local acceptance rates. Conversely, lower values sharpen the proposal, improving local accuracy at the expense of global mobility.
    \item [(2)] {\em Phase angels $\theta_{12}$ and $\theta_3$}, the rotation angles of the quantum gates,  modulate interference effects in the quantum proposal, see Fig. \ref{fig:quantum_circuit}. They influence the balance between constructive and destructive contributions in the superposition of sampling amplitudes. In practice, they serve to shape the detailed structure of the target distribution and can introduce non-trivial correlations.
    \item [(3)] {\em Parameter $p$} governs the number of quantum replicas or parallel sampling paths. Higher values generally improve convergence to the target distribution by averaging over multiple phase-shifted paths but increase computational cost.
    \item [(4)] {\em Grid bound $a_{\max}$} sets the finite support of the discretized quantum state space, i.e., the admissible acceleration samples are restricted to $\mathbf{a}\in\prod_{i\in\{x,y,z\}}\left[-a_{\max,i},\,a_{\max,i}\right]$. Increasing $a_{\max}$ enlarges the dynamic range and reduces boundary effects, but coarsens the effective resolution for a fixed qubit number $N_q$. Decreasing $a_{\max}$ improves resolution around the origin, but may truncate the tails of the target distribution, which can bias both acceptance and long-time dispersion statistics.
\end{enumerate}
In all QE-MCMC runs, we fix the grid bound based on the target acceleration covariance, $a_{\max,i}=3\sqrt{(\hat\Sigma_{\mathbf{a}\mathbf{a}})_{ii}}$ for $i\in\{x,y,z\}$. This choice provides sufficient dynamic range to capture the relevant probability mass of the target while keeping the discretization resolution adequate for the chosen qubit count.

To evaluate the influence of the remaining parameters on the accuracy of the QE-MCMC method, a parameter grid search was conducted, using the mean squared logarithmic error $\mathrm{MSLE}\,(T_{12}, T_3, \theta_{12}, \theta_3, p)$ as the objective function,
\begin{align}
\mathrm{MSLE} = \frac{1}{N_t} \sum_{n=1}^{N_t} [&\log\left(D^{\mathrm{QE-MCMC}}(t_n) + 1\right) \nonumber\\- &\log\left(D^{\mathrm{SDE}}(t_n) + 1\right)]^2.
\end{align}
This choice is motivated by the coarse discretization induced by the quantum encoding and limited qubit counts, which restrict the effective resolution of the sampler and can lead to a wide dynamic range of $D(t)$. The logarithmic transform stabilizes the optimization by reducing the dominance of late-time, large-magnitude dispersion values and by keeping the objective well-conditioned even when early-time dispersion is small.

Correlation-based calibration strategies were also explored, for instance by matching short-lag autocorrelations of the acceleration Markov chain to the target relaxation behavior of the underlying stochastic model. However, in the coupled tracer dynamics these local-in-time correlation matches did not reliably translate into agreement of the full dispersion curve across time scales. Therefore, the MSLE-based grid search is adopted as the primary calibration strategy for QE-MCMC in this work.

A systematic scan over the QE-MCMC parameter space revealed a (local) minimum of the MSLE for the following combination: $p = 10$, $T_{12} = 2.6$, $T_3 = 19.0$, $\theta_{12} = 7\pi/1024$, and $\theta_3 = 11\pi/2048$. At this configuration, the QE-MCMC method produced dispersion curves that closely matched the reference SDE solution with $\mathrm{MSLE}=0.014$, indicating that the quantum proposal mechanism, when properly tuned, can effectively replicate the correct statistical behavior of the underlying dynamics, see again Fig. \ref{fig:dispersion_curves_shear_flow}.

\subsubsection{Spectral gap analysis}
To quantitatively assess the potential quantum advantage, we compared the spectral gaps $\delta$ of the Markov matrices of the classical MCMC and quantum-enhanced MCMC algorithms. Since the spectral gap governs the asymptotic convergence rate of a Markov chain in the classical sense (see Sec. I), larger values of $\delta$ are typically associated with faster mixing and thus more efficient sampling. In the present framework, however, this interpretation also carries over in a more structured way, as the Markov chain is calibrated such that its temporal correlations are consistent with the physically prescribed relaxation dynamics. Consequently, the spectral gap remains linked to an effective convergence rate, but only insofar as it preserves the target correlation time scale of the underlying stochastic process. 

\begin{figure}[t]
    \centering
    \includegraphics[width=.48\textwidth]{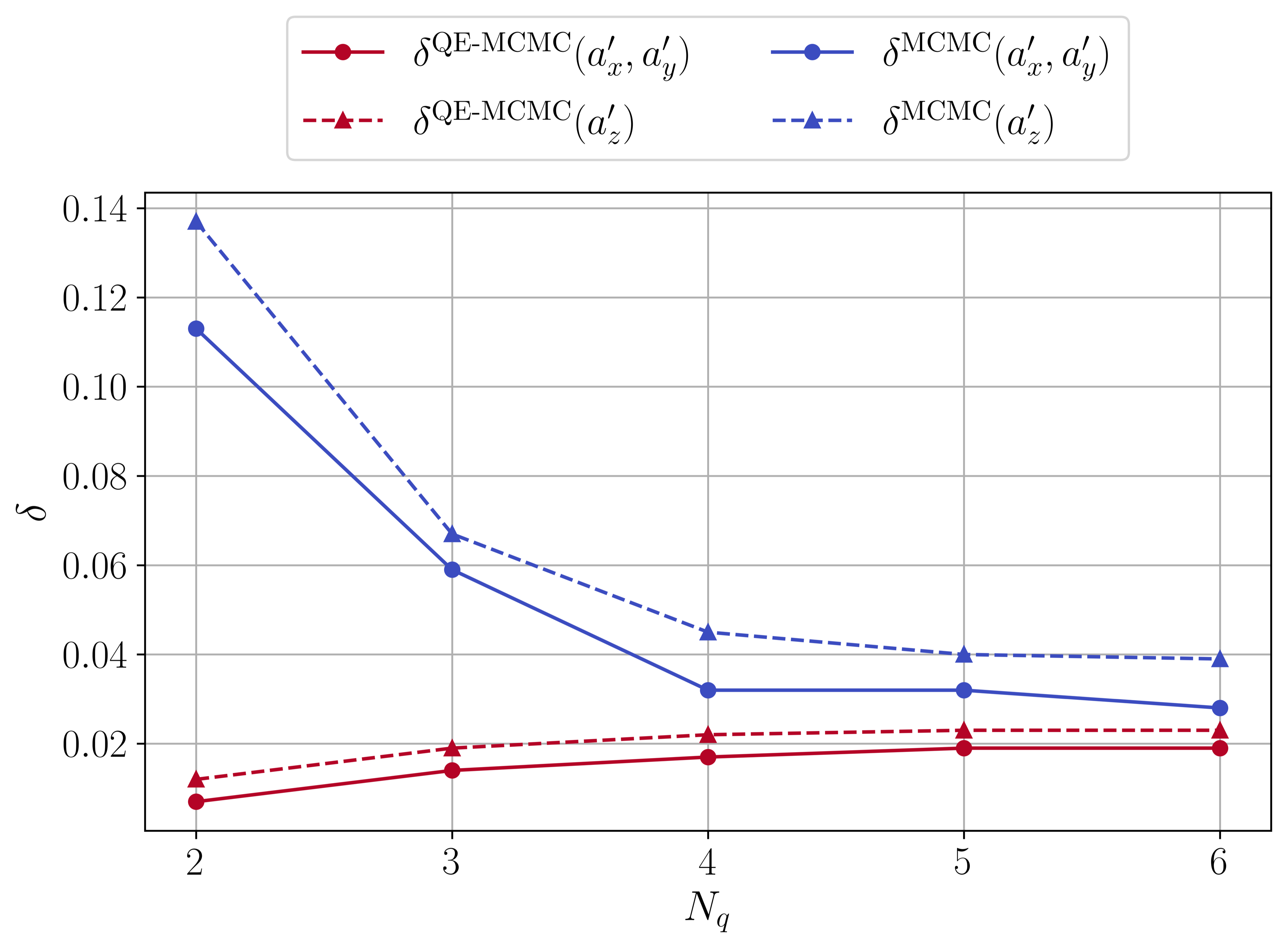}
    \caption{Comparison of the spectral gap $\delta$ between classical MCMC and QE-MCMC as a function of the number of qubits per dimension $N_q$ for the HSF case. Here, the number of layers is $p=10$ in the quantum algorithm.}
    \label{fig:delta_vs_N}
\end{figure}

In the QE-MCMC case, the transition matrix naturally has finite dimension of
\begin{equation}
    \dim(\Pi) = 2^{d\times N_q}\,, 
\end{equation}
where $d$ denotes the dimensionality of the distribution, such as $d=1$ for $a_z$ and $d=2$ for $a_{x}$ and $a_y$. Furthermore, $N_q$ is the number of qubits per dimension $d$. In contrast, the classical MCMC operates in a continuous state space, implying an infinite-dimensional transition operator. To enable a consistent comparison, we discretized the MCMC samples by grouping them into the same effective resolution as in the QE-MCMC method, thereby constructing transition matrices of identical dimension. The resulting comparison is shown in Fig. \ref{fig:delta_vs_N}. It is seen that the spectral gap for classical sampling algorithm is always larger than the one for the quantum-enhanced algorithm, which would imply that no quantum advantage is obtainable. However, we observe that the difference between classical and quantum case is steadily reduced when the qubit number and thus the resolution of the sampling distribution is enhanced. We note that the QE-MCMC case operates at sampling temperatures $T>2$, which results in a tempered target PDF.

\subsection{Turbulent channel flow}

\subsubsection{Flow configuration}
\begin{figure}
    \centering
    \includegraphics[width=0.48\textwidth]{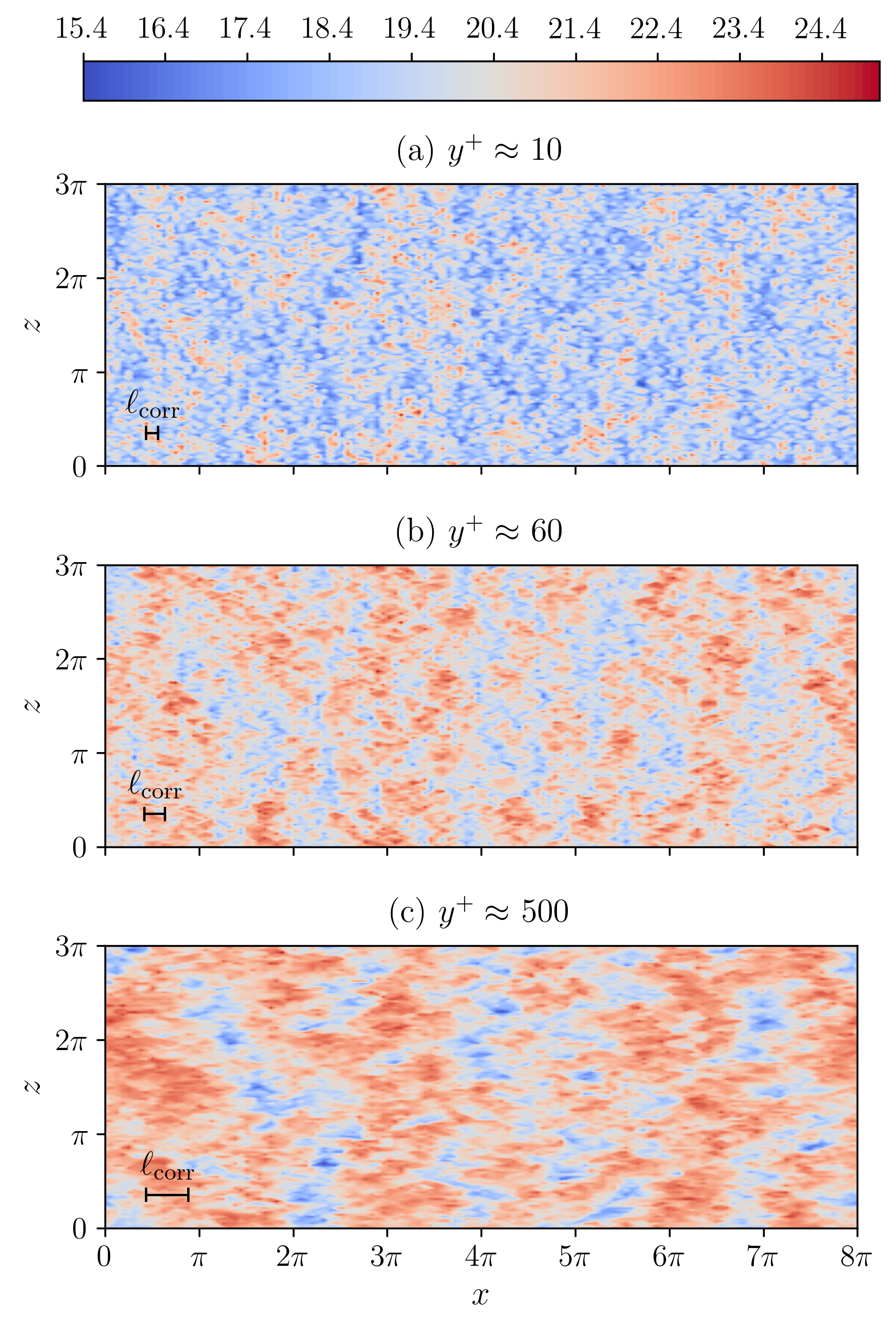}
    \caption{Contours of the streamwise velocity component $u_x^+(x^+, y^+, z^+, t^+\approx250)$ at a time instant. The horizontal slice cuts are taken at three distances from the wall, (a) in the viscous sublayer at $y^+\approx 10$, (b) in the logarithmic region at $y^+\approx 60$, and (c) in the center region at $y^+\approx 500$. The midplane of the channel is at $y=h$, which corresponds to $y^+=1000$. Data are taken from the Johns Hopkins Turbulence Database \cite{Graham2016}. In each panel, we indicate the correlation scale $\ell_{\rm corr}$, which increases with distance from the wall.}
    \label{fig:ux_over_xz}
\end{figure}

The second application case is a turbulent channel flow (TCF), a three-dimensional Navier-Stokes flow between two parallel plates of distance $2h$ that is driven by a constant pressure gradient $\mathrm{d}p/\mathrm{d}x$ in streamwise direction $x$ \cite{Kim1987,Pope2000}. In this configuration, two turbulent boundary layers form at the bottom and top plates, which grow up from both walls to the mid-plane of the channel (where they meet). Since there is a statistical top-down symmetry, we will consider the lower half channel only for our study, $0\le z\le h$. The Lagrangian particle dispersion will be studied in the lower channel half for the following. In the wall-normal direction, at $y=\{0,2h\}$ the fluid satisfies the no-slip boundary condition, $\mathbf{u}=0$; in streamwise ($x$) and spanwise ($z$) periodic boundary conditions apply. The velocity field $\mathbf u$ is obtained by direct numerical simulations (DNS) of the incompressible Navier-Stokes equations, which are given by 
\begin{align}
\frac{\partial \mathbf u}{\partial t}+(\mathbf u\cdot \boldsymbol \nabla)\mathbf u &=-\frac{1}{\rho_0}\boldsymbol \nabla P + \nu\boldsymbol\nabla^2\mathbf u\,\\
\boldsymbol \nabla\cdot \mathbf u &=0\,,
\end{align}
with constant mass density of the fluid $\rho_0$, pressure field $P$, and kinematic viscosity $\nu$.

\begin{figure*}[t]
    \centering
    \includegraphics[width=0.95\textwidth]{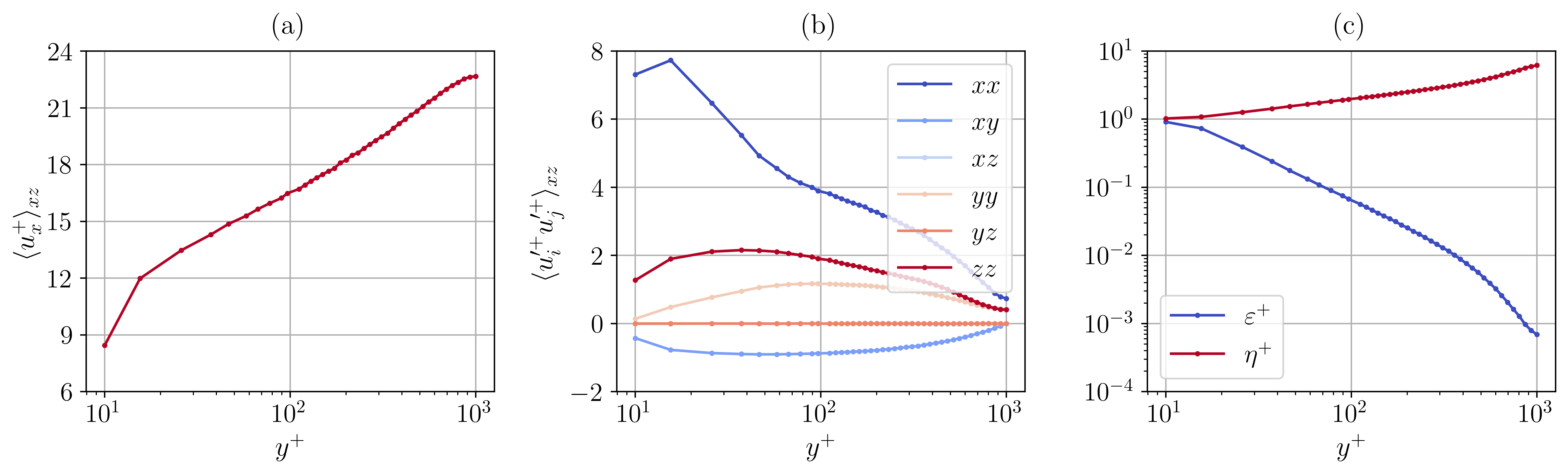}
    \caption{Wall-normal mean profiles of essential turbulent properties plotted in wall units for 64 selected horizontal planes, that will be used for the models, in semi-logarithmic plots. (a) Mean streamwise velocity $\langle u_x^+\rangle_{xz}(y^+)$. (b) Components of the symmetric Reynolds stress tensor $\langle u_i^{\prime +}u_j^{\prime +}\rangle_{xz}(y^+)$. (c) Plane-averaged mean Kolmogorov scale $\eta^+(y^+)=\eta(y)/\ell^+$ and mean kinetic energy dissipation rate $\varepsilon^+(y^+)=\varepsilon(y)\ell^+/u_{\tau}^3$. Note that the vertical grid resolution of the DNS is given with 512 non-uniform grid points, much finer than the selected 64 points, which are shown here.}
    \label{fig:profiles}
\end{figure*}

Figure \ref{fig:ux_over_xz} shows three horizontal contour cuts at a time instant through the channel volume $V=L_x\times 2h \times L_z=8\pi h \times 2h \times 3\pi h$ taken at different distances from the bottom wall. The distance is measured in wall or inner units, which are given by $\ell^+=\nu/u_{\tau}$ with $u_{\tau} = \sqrt{\tau_w/\rho_0}$. Here, $u_{\tau}$ is the friction velocity, $\tau_w$ the wall shear stress --- the viscous stress at the wall. The coordinates and the velocity field components are expressed in wall units: $x_i^+=x_i/\ell^+$ and $u_i^+=u_i/u_\tau$. Consequently, a dimensionless time $t^+=t u_\tau/\ell^+$ follows. We observe that the structures of the streamwise velocity component $u_x^+$ display contours that are increasingly elongated in the streamwise direction. We quantify the streamwise correlation length in the TCF case by
\begin{equation}
    \ell_{\rm corr}(y^+) = \int_{0}^{\infty} C_{11}(r,y^+)\,\mathrm{d}r,
\end{equation}
where the normalized streamwise autocorrelation of the streamwise velocity fluctuations is
\begin{equation}
    C_{11}(r,y^+) = \frac{\left\langle u_x^\prime(x,z,t;y^+)\,u_x^\prime(x+r,z,t;y^+)\right\rangle_{x,z,t}}
    {\left\langle u_x^{\prime\,2}(x,z,t;y^+)\right\rangle_{x,z,t}}.
\end{equation}
The DNS turbulence data, which we will use to build the target distributions, are at a friction Reynolds number \cite{Graham2016}
\begin{equation*}
Re_{\tau}= \frac{u_\tau h}{\nu}=1000\,.
\end{equation*}
Figure \ref{fig:profiles} replots the essential statistical quantities for the MCMC and QE-MCMC models versus $y^+$. These are the mean velocity in panel (a), the Reynolds stress components in panel (b), and the Kolmogorov scale $\eta(y)=\langle\nu^3/\varepsilon(y)\rangle_{xz}^{1/4}$, which quantifies the size of the smallest vortical structures at distance $y$, and the kinetic energy dissipation rate $\varepsilon(y)=\nu \langle (\boldmath\nabla \mathbf u)^2\rangle_{xz}$ in panel (c). The averages are combined time- and $xz$-plane-averages which are given by (here for a single velocity component)
\begin{equation}
\langle u_i^+\rangle_{xz}(y^+) = \frac{1}{L^+_x L^+_z}\int_0^{L^+_x}\int_0^{L^+_z} \langle u_i^+\rangle\, \mathrm dz^+\, \mathrm dx^+ 
\end{equation}
with
\begin{equation}
\langle u_i^+\rangle = \frac{1}{T}\int_0^T u_i^+ \mathrm dt\,. 
\end{equation}
Here, $T$ is the total time interval. The linear scaling in panel (a) for $30\lesssim y^+\lesssim 200$ defines the so-called logarithmic region of the channel flow. Unlike the homogeneous shear flow benchmark, the statistical quantities are now height-dependent. The turbulence is not homogeneous with respect to the wall-normal direction, but remains homogeneous with respect to the streamwise and spanwise directions. This adds a further dependence to the probability density functions, namely the height dependence, which increases the complexity of the modeling task significantly.  

\begin{figure*}[ht]
    \centering
    \includegraphics[width=0.97\linewidth]{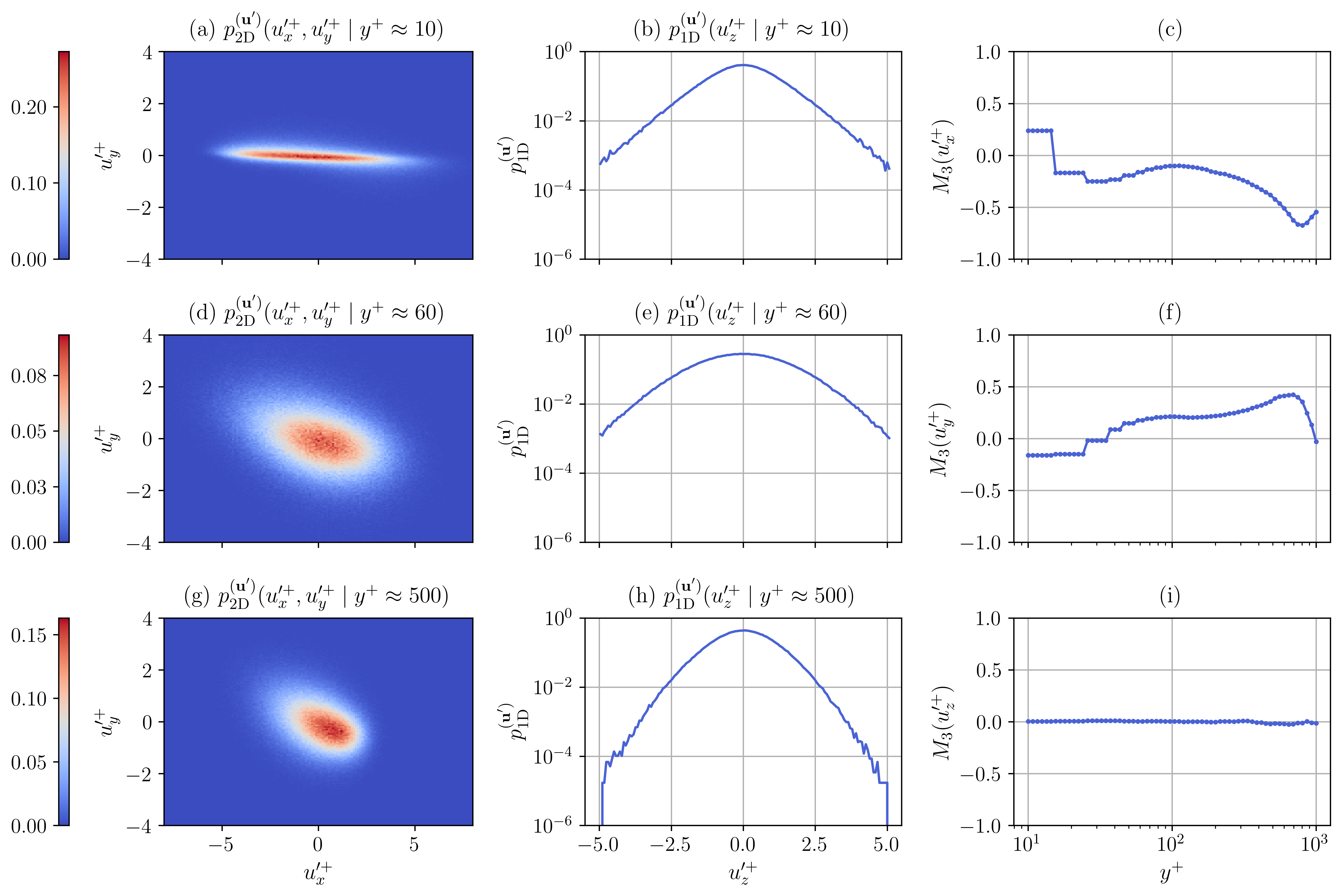}
    \caption{Statistics of the velocity fluctuations in the turbulent channel flow at different heights. Panels (a,b) display data for $y^+\approx 10$, (d,e) for $y^+\approx 60$, and (g,h) for $y^+\approx 500$. Panels (a,d,g) show the joint conditional PDFs $p_{\rm 2D}^{(\mathbf{u}^\prime)}(u_x^{\prime +},u_y^{\prime +}|\,y^+)$. Panels (b,e,h) the spanwise PDFs $p_{\rm 1D}^{(\mathbf{u}^\prime)}(u_z^{\prime +}|\,y^+)$. Finally, panels (c), (f), and (i) plot the skewness $M_3$ of the velocity $u_x^{\prime +}$, $u_y^{\prime +}$, and $u_z^{\prime +}$, respectively, versus wall normal coordinate $y^+$ at the selected 64 analysis planes.}
    \label{fig:channelstat}
\end{figure*}

\subsubsection{Height dependence of velocity fluctuation statistics}
We proceed similarly to the HSF case. The acceleration statistics will be inferred from the velocity fluctuation statistics as discussed in subsection III.A. Complexity is, however, significantly increased by the dependence of the turbulent fluctuations from the distance to the wall.

Figure \ref{fig:profiles} shows that the Reynolds stresses $\langle u^{\prime +}_i u^{\prime +}_j\rangle_{xz} =0$ for the cross correlations $(i,j)=(x,z)$ and $(i,j)=(y,z)$, which suggests a factorization similar to the homogeneous shear flow case:
\begin{align}
p^{(\mathbf{u}^\prime)}(u_x^{\prime +},u_y^{\prime +},u_z^{\prime +}|\,y^+)=\,&p^{(\mathbf{u}^\prime)}_{\rm 2D}(u_x^{\prime +},u_y^{\prime +}|\,y^+)\,\otimes \nonumber\\
&p^{(\mathbf{u}^\prime)}_{\rm 1D}(u_z^{\prime +}|\,y^+)\,.
\label{tensorTCF}
\end{align}
Compare the probability density in Eq. \eqref{tensorTCF}, that now contains a height-dependence, with Eq. \eqref{tensorHSF}. We do not display data in the viscous sublayer, $y^+\lesssim 10$. In this near-wall region the velocity-fluctuation distributions extracted from the database are strongly non-Gaussian and exhibit pronounced skewness and intermittency, such that a skew-normal approximation does not provide a reliable fit.

Figure \ref{fig:channelstat} discusses the PDFs of the velocity fluctuations at different heights $y^+$. Panels (a), (d), (e) display the JPDFs $p_{\rm 2D}^{(\mathbf{u}^\prime)}(u_x^{\prime +},u_y^{\prime +}|\,y^+)$. They are skewed with respect to both velocity components as shown in corresponding panels (c), (f), (i) where we calculate the skewness $M_3(u_i^{\prime +})=\langle (u_i^{\prime +})^3\rangle_{xz}/\langle (u_i^{\prime +})^2\rangle_{xz}^{3/2}$ versus $y^+$ for the 64 selected planes of the lower half channel. We also recognize that the distribution of the spanwise velocity fluctuation component $p_{\rm 1D}^{(\mathbf{u}^\prime)}(u_z^{\prime +}|\,y^+)$ in panel (h) is symmetric with $M_3$ almost perfectly zero for all $y^+$, which is visible in panel (i). Also observable is a change of the sign of the skewness of the streamwise and wall-normal components; while the streamwise velocity component switches from a fatter positive to a negative tail at the beginning of the logarithmic layer ($y^+\simeq 30$), the wall-normal component shows a reverse trend. This statistical behavior is connected to the formation of streamwise streaks by counter-rotating streamwise vortices in the presence of height-dependent mean shear.        

\begin{figure*}[ht]
    \centering
    \includegraphics[width=0.95\textwidth]{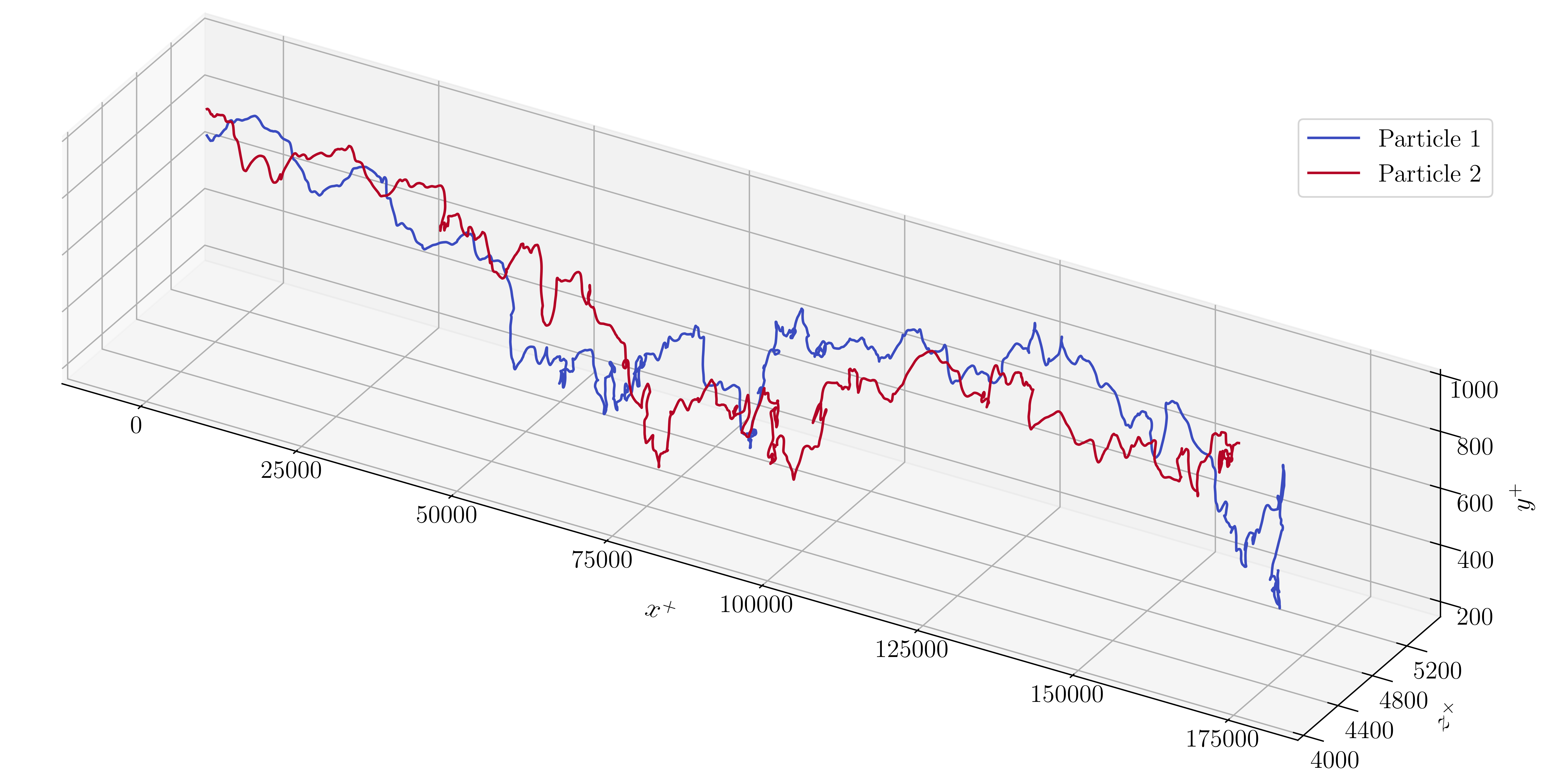}
    \caption{Temporal evolution of two Lagrangian particles within a tracer pair over $10^5$ steps which were generated by the QE-MCMC algorithm. They were initialized at $\mathbf{X}^{+}(0)=[0, 500, L_z^+/2]^\mathrm{T}$ and $\tilde{\mathbf{X}}^{+}(0)=[0, 500+16\eta^+, L_z^+/2]^\mathrm{T}$, respectively. The initial fluctuating velocities and accelerations were sampled as $(\mathbf{u}^{\prime}(0),\mathbf{a}(0))^\mathrm{T}\sim\mathcal{N}(\mathbf{0},\hat{\Sigma}_{\mathbf{z}\mathbf{z}}(y_0^+))$.}
    \label{fig:trajectories_3d}
\end{figure*} 

Figure \ref{fig:channelstat} also underscores that the factorized distributions can be approximated as skew normal distributions, which are given by (also cf. Eq. \eqref{eq:Pxy_HSF_u} for the homogeneous shear flow)
\begin{align}
p^{(\mathbf{u}^\prime)}_{\rm 2D}&(u_x^{\prime +}, u_y^{\prime +}|y^+) = \nonumber\\
&\frac{1}{2\pi\,|\hat{\Sigma}_{\rm 2D}^{(\mathbf{u}^\prime)}|^{1/2}} \exp\left( -\frac{1}{2} (\mathbf{u}^{\prime +})^\mathrm{T} \bigl(\hat\Sigma_{\rm 2D}^{(\mathbf{u}^\prime)}\bigr)^{-1} \mathbf{u}^{\prime +} \right) \times \nonumber\\
& 2 \Phi (\sqrt{2}\boldsymbol{\kappa}^{\rm T}(\hat\Sigma_{\rm 2D}^{(\mathbf{u}^\prime)})^{-1/2} \mathbf{u}^{\prime +})
\label{eq:Pxy_TCF}
\end{align}
with $\Phi(x)=[1+{\rm erf}(x/\sqrt{2})]/2$ and 
\begin{align} 
\mathbf{u}^{\prime +} = \begin{bmatrix} u_x^{\prime +} \\ u_y^{\prime +} \end{bmatrix}, \, \hat\Sigma_{\rm 2D}^{(\mathbf{u}^\prime)}(y^+) = \begin{bmatrix} \Sigma_{xx}(y^+) & \Sigma_{xy}(y^+) \\ \Sigma_{xy}(y^+) & \Sigma_{yy}(y^+) \end{bmatrix}.
\end{align}
Similarly, the target distribution for the spanwise component follows to 
\begin{align}
p_{\rm 1D}^{(\mathbf{u}^\prime)}(u_z^{\prime +}|\,y^+) =& \frac{1}{\sqrt{2\pi\, \sigma_u^2}} \exp\left( -\frac{u_z^{\prime +\, 2}}{2\sigma_u^2} \right) \times \nonumber\\ & 2\Phi(\sqrt{2}\kappa_z\sigma_u^{-1}u_z^{\prime +})  
\label{eq:Pz_TCF}
\end{align}
with $\sigma_u^2(y^+) = \Sigma_{zz}(y^+)$. The parameter vector $\boldsymbol\kappa(y^+)=[\kappa_x(y^+),\kappa_y(y^+),\kappa_z(y^+)]^\mathrm{T}$ captures the change of the sign, $\kappa_i(y^+)=\textrm{sgn}[M_3(u_i^{\prime +},y^+)]$ in Eqs. \eqref{eq:Pxy_TCF} and \eqref{eq:Pz_TCF}. It is expected that the skewness for the spanwise statistics is significantly less skewed as for the other two space directions, which are correlated. It could be well approximated by a normal distribution in the core of the fluctuation range. We will stick to the generalized version of $p_{\rm 1D}(u_z')$, similar to the other two components. 
As mentioned in the HSF case, the velocity statistics enter our stochastic two-equation model through the diffusion tensor $G_{ij}(y^+)$, which is obtained from the velocity covariance by solving the stationary Lyapunov equation (see again the appendix). This yields a height-dependent acceleration covariance $\hat\Sigma_{\mathbf{a}\mathbf{a}}(y^+)=\frac{1}{2}\tau_\eta \hat G(y^+) \hat G^\mathrm{T}(y^+)$ and thus the corresponding conditional acceleration target distributions $p_{\rm 2D}^{(\mathbf{a})}(a_x^+,a_y^+\,|\,y^+)$ and $p_{\rm 1D}^{(\mathbf{a})}(a_z^+\,|\,y^+)$ used by MCMC and QE-MCMC.

\begin{figure*}[ht]
    \centering
    \includegraphics[width=0.8\textwidth]{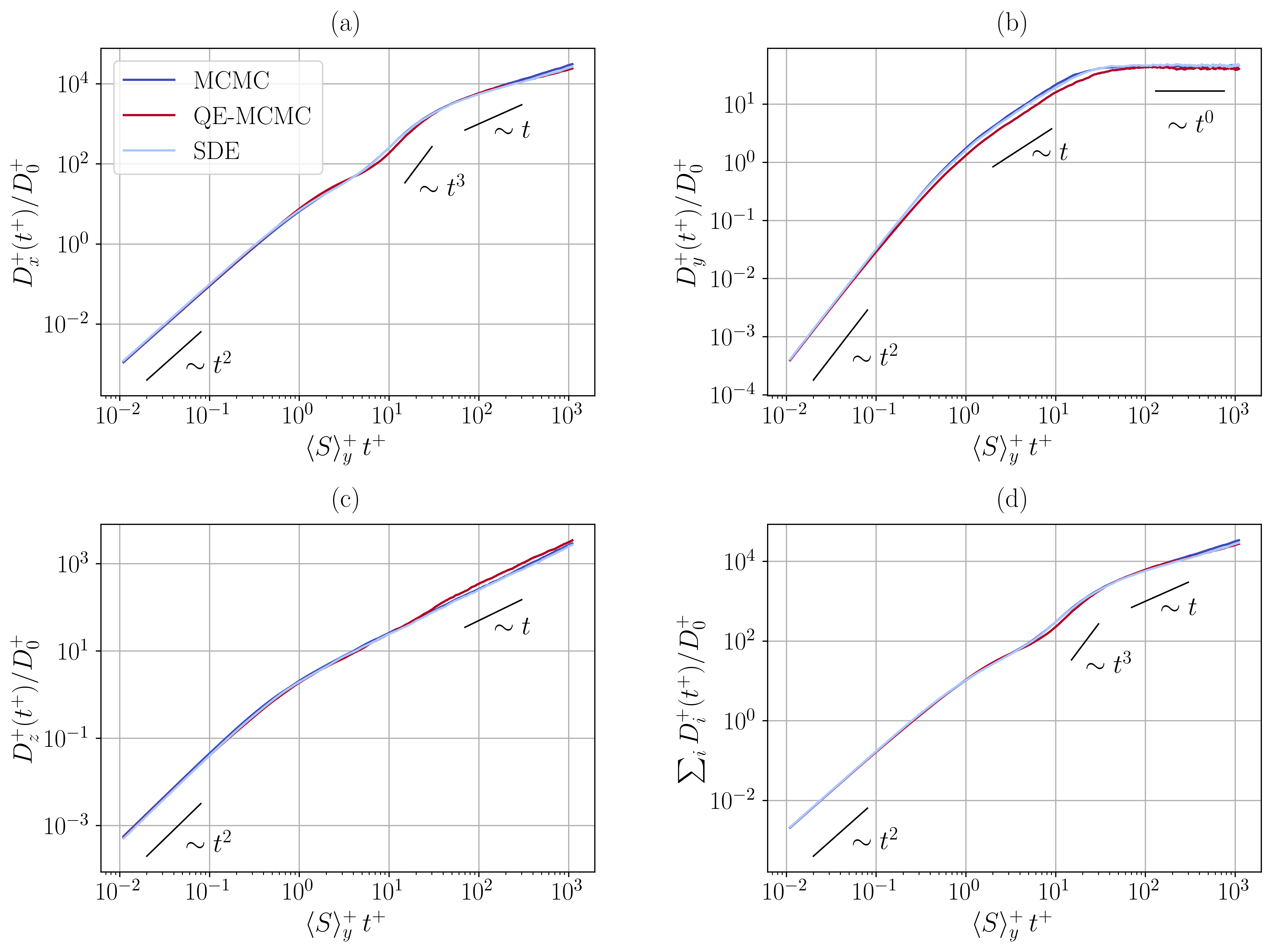}
    \caption{Lagrangian particle pair dispersion curves. We compare a classical stochastic model (SDE) with a classical and quantum-enhanced Markov chain Monte Carlo method. We seed 3000 Lagrangian tracer particles in pairs at initial positions $\mathbf{X}^{(k)+}(0)=[0, 500, L_z^+/2]^\mathrm{T}$, $\tilde{\mathbf{X}}^{(k)+}(0)=[0, 500+16\eta^+, L_z^+/2]^\mathrm{T}$ and sampled initial velocity and acceleration $(\mathbf{u}^{\prime\,(k)}(0),\mathbf{a}^{(k)}(0))^\mathrm{T}\sim\mathcal{N}(\mathbf{0},\hat{\Sigma}_{\mathbf{z}\mathbf{z}}(y_0^+))$.}
    \label{fig:dispersion_curves_channel_flow}
\end{figure*} 

\subsubsection{Lagrangian particle pair dispersion}
Similarly to subsection III.A.2, we analyze the Lagrangian particle pair dispersion. The definitions \eqref{LagDisp} are used again. Figure \ref{fig:trajectories_3d} displays two trajectories within a tracer pair, that were initialized close to each other. It is observed that they stay together for a period before separating. This is caused by spatially extended streamwise velocity fluctuations, which are observable in the contours in Fig. \ref{fig:ux_over_xz}.     

The number of qubits per dimension in the QE-MCMC case is mostly $N_q=5$; the statistics is collected again for $N_{\rm pa}=3000$. We need a volume measure of the mean shear. Thus, we first define the following mean derivative 
\begin{equation}
S_h^+(y^+) = \frac{\mathrm{d}\langle u\rangle_{xz}^+}{\mathrm{d}y^+} \;\; \mbox{and}\;\;
\langle S\rangle^+_y =\frac{1}{L_y^+}\int_0^{L_y^+}S_h^+(y^+)\,\mathrm{d}y^+\,.
\end{equation}
The global mean shear time $T_{S^+}$ follows to 
\begin{equation}
T_{S^+}= \frac{1}{\langle S\rangle_y^+}\,.
\label{meanshear}
\end{equation}
For the present data record, we obtain $T_{S^+}\approx 91$ time units. 
In this analysis, we enforce reflective (specular) boundary conditions at $y^+=20$ and $y^+=1000$ via ideal elastic wall collisions, i.e.\ any boundary overshoot is corrected by mirroring the wall-normal position back into the domain and reversing the wall-normal components, $u_y^{+}\!\to -u_y^{+}$ and $a_y^{+}\!\to -a_y^{+}$.
As already stated, for $y^+<20$ the velocity statistics deviates strongly from assumptions that we made. We note here that we cannot provide a comparison with Lagrangian trajectory data obtained from the DNS data since the snapshot output interval is too coarse for a particle advection.  

The results are shown in Fig. \ref{fig:dispersion_curves_channel_flow} for the individual space directions and their sum. All curves start with a ballistic scaling $D_i(t)\sim t^2$ for small time scales $t^+\ll T_{S^+}$ with the mean shear-based characteristic time scale $1/\langle S\rangle_y^+$. For times $t^+\gtrsim T_{S^+}$, we either observe a crossover to a diffusive scaling $D_i(t)\sim t$ for the wall-normal and spanwise directions, or a superdiffusive scaling of $D_i(t)\sim t^3$ for the streamwise direction (and thus for the total dispersion). This scaling behavior is similar to the HSF benchmark case, where such an intermediate superdiffusive scaling was observed for all space directions. We suspect that the absence of a superdiffusive dispersion for wall-normal and spanwise directions is caused by the shear rate dependence on $y^+$. The present TCF dispersion behavior has also been observed in previous DNS studies, e.g. by Polanco et al. \cite{Polanco2018}. Note that for the present case the mean shear rate depends on the distance form the wall (or ground); the shear scale $T_{S^+}$ is an average measure as given in Eq. \eqref{meanshear}. Eventually, for $t^+\gg T_{S^+}$, we obtain a diffusive scaling in streamwise and wall-normal directions. The level-off for $D_y(t)$ is caused by the constraining of the vertical motion in the lower half channel. We again find very good agreement in the pair-dispersion statistics for all three methods.

A cubic scaling of the pair dispersion is otherwise obtained from the classical Kolmogorov-Obukhov (K41) theory of the inertial cascade range of turbulence \cite{Obukhov1941}, see also \cite{Boffetta2002,Birnir2025}. Starting point is 
\begin{equation}
\frac{\mathrm{d}D}{\mathrm{d}t}=\tau(\ell) \langle(\delta_{\ell}u)^2\rangle\,,
\label{eq:Obukhov}
\end{equation}
with an eddy turnover time scale at scale $\ell$ of $\tau(\ell)=\varepsilon^{-1/3}\ell^{2/3}$ and a second-order velocity increment moment of $\langle(\delta_{\ell}u)^2\rangle=C_2\varepsilon^{2/3}\ell^{2/3}$. This leads to $\mathrm{d}D/\mathrm{d}t=C_2\varepsilon^{1/3}\ell^{4/3}$.     Inserting $\ell=\sqrt{D}$ into Eq.~\eqref{eq:Obukhov} yields 
\begin{equation}
D(t)=\frac{C_2}{27} \varepsilon t^3\,.
\label{eq:Obukhov1}
\end{equation}
The constant $C_2\approx 2$, such that the prefactor $g^{\rm K41}=C_2/27\approx 0.07$ \cite{Mazzitelli2014}. Here, we obtain $g\simeq 0.13$. We mention that such an intermediate scaling $D\sim t^{\alpha}$ with $\alpha\simeq 3$ to 4, as seen here for HSF and TCF, is observed in other flows, such as turbulent convection \cite{Schumacher2008,Ettel2026}.    

\begin{figure*}[ht]
    \centering
    \includegraphics[width=0.9\textwidth]{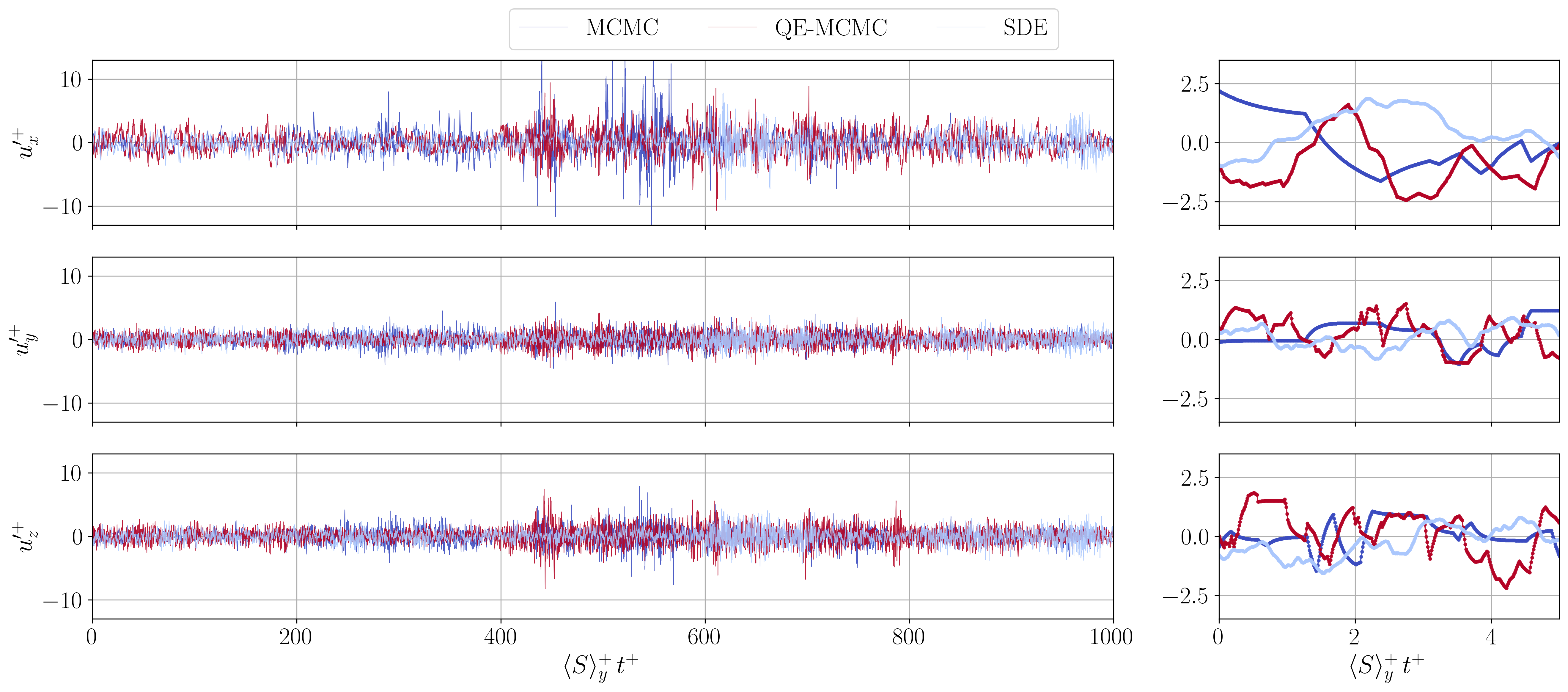}
    \caption{Time traces of the three components of the velocity fluctuation vector field, $u_i^{\prime +}$. We compare the solution of the stochastic differential equation (SDE), with the classical and quantum-enhanced Markov chain Monte Carlo method, (MCMC) and  (QE-MCMC). The left column shows the time traces for 1000 time units. The right column displays a zoom. The legend at the top holds for all panels.}
    \label{fig:traceplot}
\end{figure*} 

Figure \ref{fig:traceplot} compares the time series of the three velocity fluctuation components, generated by the three algorithms that are compared with each other. A zoom to the right of each panel shows that the time series generated by QE-MCMC method are comparable the corresponding classical counterparts. We observed less rejections in the QE-MCMC case than in the MCMC case, which generated slightly smoother time series. We can conclude that a small number of qubits can generate synthetic time traces effectively.

\begin{figure*}[ht]
    \centering
    \includegraphics[width=0.99\textwidth]{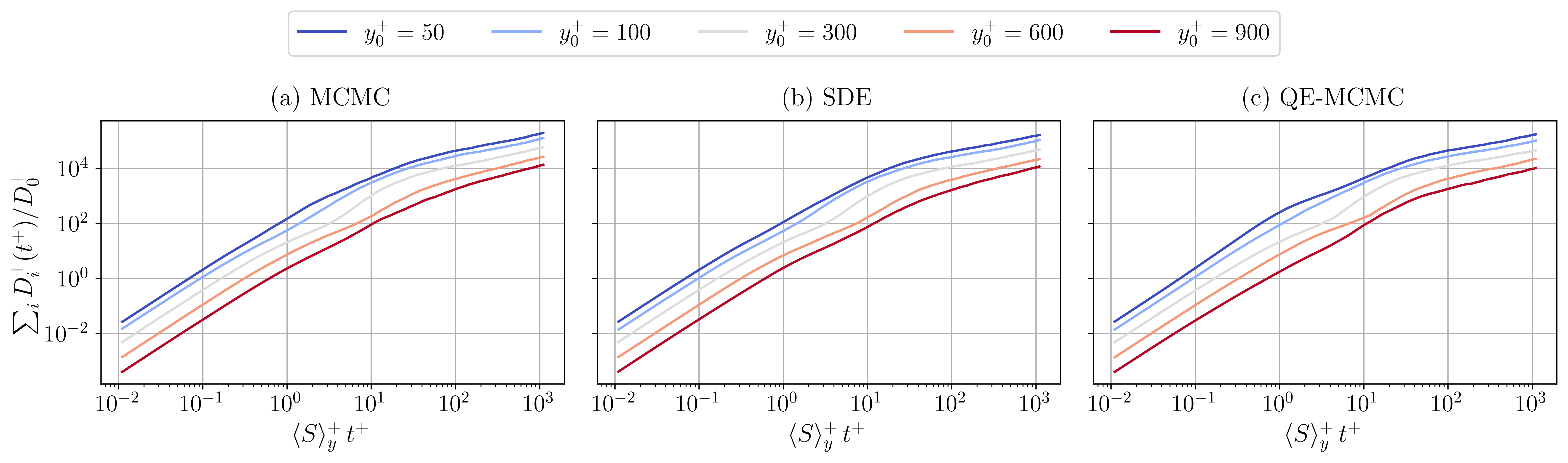}
    \caption{Total particle pair dispersion curves for different initial seeding heights, which are indicated in the legend on top of the panels. (a) Classical MCMC algorithm. (b) Stochastic model. (c) Quantum-enhanced MCMC algorithm. The temperature parameter $T$ was set to 3.4 for the two-dimensional and 3.9 for the one-dimensional distribution.}
    \label{fig:dispersion_curves_over_y0}
\end{figure*} 

Next, we investigate the height dependence of the total dispersion. Figure \ref{fig:dispersion_curves_over_y0} compares the results for the three methods and 5 different seeding positions of the tracer pairs, as indicated in the legend of the figure. All curves start with a ballistic scaling $\sim t^2$. A diffusive range with a $t$--scaling range is found for all data at the largest times. For intermediate seeding heights beyond the logarithmic region a crossover to a short superdiffusive scaling is detected, consistent with the findings in Fig. \ref{fig:dispersion_curves_channel_flow}. It is also seen that all three methods give again nearly the same results.  

\subsubsection{Optimal parameters of the MCMC and QE-MCMC method}
For the turbulent channel flow case, we employ the same calibration strategies as in the HSF benchmark to determine the sampling parameters of both MCMC and QE-MCMC. Due to the height-dependent flow statistics, the classical MCMC proposal scales must be adapted locally in $y^+$, i.e.\ $\sigma_{12,\mathrm{prop}}=\sigma_{12,\mathrm{prop}}(y^+)$ and $\sigma_{3,\mathrm{prop}}=\sigma_{3,\mathrm{prop}}(y^+)$, in order to reproduce the prescribed height-dependent target statistics and the resulting dispersion behavior. In the QE-MCMC method, the finite quantum grid bounds are likewise chosen height-dependent via $a_{\max,i}(y^+)=3\sqrt{(\hat\Sigma_{\mathbf{aa}})_{ii}(y^+)}$ to accommodate the changing acceleration variance across the channel. In contrast, the remaining QE-MCMC circuit parameters can be kept constant over the height, i.e.\ $(\theta_{12},\theta_3,T_{12},T_3)$ are selected globally, which indicates a stronger degree of parameter transferability (generalization) of the quantum proposal mechanism compared to the classical random-walk proposal. For the turbulent channel flow considered here, the resulting optimal QE-MCMC parameters are $\theta_{12}=8\pi/1024$, $\theta_3=33\pi/2048$, $T_{12}=3.4$, and $T_{3}=3.9$. We used $p=10$ as in the shear-flow case.

\subsubsection{Spectral gap analysis}
Due to the wall-normal inhomogeneity of the turbulent channel flow, the acceleration statistics and thus the Markov transition mechanism depend on the instantaneous wall distance. Consequently, we consider a height-conditioned transition probability $\Pi(y^+)$ and define the local spectral gap as $\delta_y(y^+) = 1 - |\lambda_2(\Pi(y^+))|$, where $\lambda_2$ denotes the second-largest eigenvalue of $\Pi(y^+)$ in magnitude. In order to obtain a single scalar measure that is representative for the overall tracer dynamics, we introduce an {\em effective spectral gap width} by averaging $\delta_y(y^+)$ with the wall-normal occupancy probability density $p_Y(y^+)$,
\begin{equation}
    \delta = \int p_Y(y^+)\,\delta_y(y^+)\,\mathrm{d}y^+.
    \label{effgap1}
\end{equation}
This weighted averaging accounts for the fact that different wall-normal regions contribute unequally to the overall tracer dynamics.
Here, $p_Y(y^+)$ is defined as the long-time empirical distribution of the tracer wall-normal coordinate $Y^+(t)$,
\begin{equation}
    p_Y(y^+) = \lim_{T\to\infty}\frac{1}{T}\int_0^T \delta_{\mathrm{D}}\!\bigl(y^+-Y^+(t)\bigr)\,\mathrm{d}t,
    \label{effgap2}
\end{equation}
with $\delta_{\mathrm{D}}$ denoting the Dirac delta distribution. This definition ensures that $\delta$ reflects the turbulent mixing performance in those wall-normal regions correctly that are actually visited by the tracers. Clearly, Eqns.~\eqref{effgap1} and~\eqref{effgap2} define a heuristic effective measure of the spectral gap. This indicator is applied consistently to both MCMC algorithms, enabling a direct comparison, however in a less rigorous sense than the original spectral gap.

\begin{figure}[t]
    \centering
    \includegraphics[width=.48\textwidth]{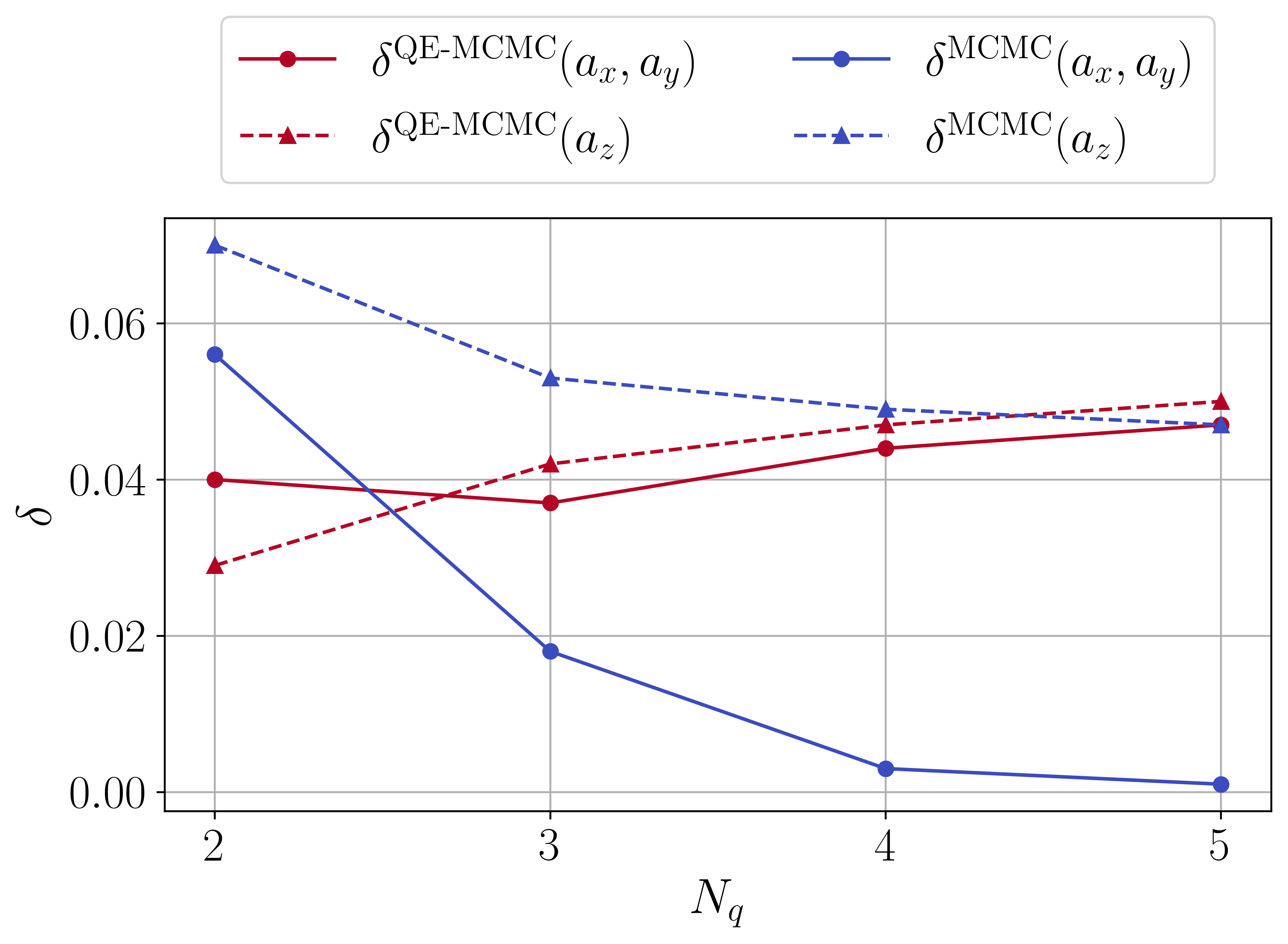}
    \caption{Comparison of the effective spectral gap $\delta$, as defined in Eq.~\eqref{effgap1}, between classical MCMC and QE-MCMC as a function of the number of qubits per dimension $N_q$ for the TCF case. Here, the number of layers is $p=10$ in the quantum algorithm.}
    \label{fig:delta_vs_N1}
\end{figure}

The comparison of the spectral gap for the classical and quantum-enhanced algorithm for qubit numbers per dimension between 2 and 5 is reported in Fig. \ref{fig:delta_vs_N1}. Similar to the subsection in III.A, both methods use the same coarse grain level for a fair comparison. First, we find that the analysis is relatively insensitive with respect to the number of layers; here we take $p=10$, which suggests that the expressivity cannot be further increased by enhancing the depth of the parametric circuit. Similar observations have been made in other applications cases, such as the solution of linear and nonlinear partial differential equations with variational quantum algorithms \cite{Gubaev2025,Koecher2025}. 

The differences between classical and quantum algorithms become now significant for the highest resolutions as seen in Fig. \ref{fig:delta_vs_N1}. Differently to the HSF case, the most complex sampling task, the bivariate PDF at highest resolutions clearly favors QE-MCMC for $N_q\ge 3$, see again Fig. \ref{fig:delta_vs_N}. The one-dimensional distribution tasks, in contrast, exhibit comparable spectral gaps for both methods.

\begin{figure}[t]
    \centering
    \includegraphics[width=.45\textwidth]{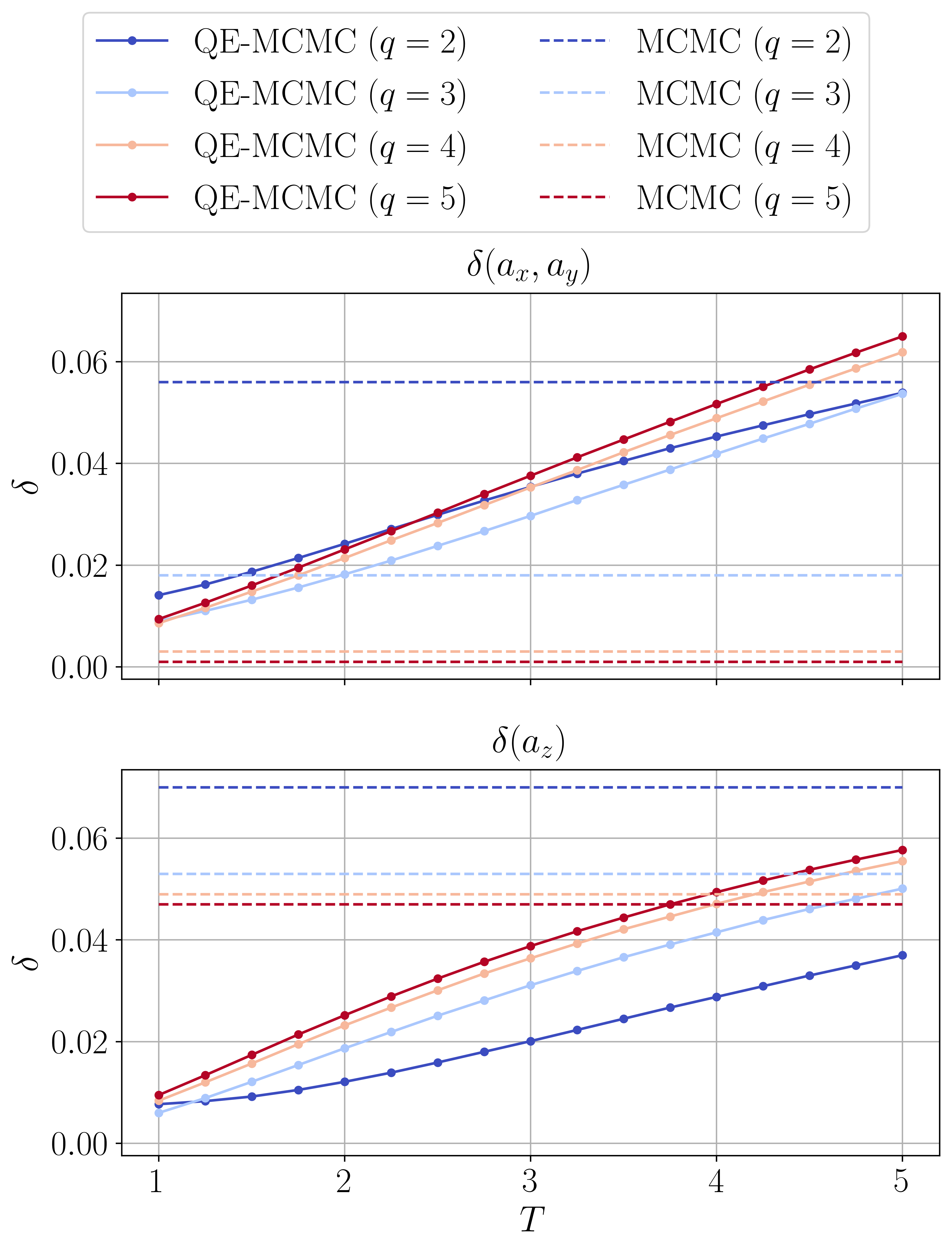}
    \caption{Dependence of the effective spectral gap width on the sampling temperature $T$ for a given qubit number $2\le N_q\le 5$. Top: Data for the joint PDF of $a'_x$ and $a'_y$. Bottom: Data for the PDF of $a'_z$. In both panels, the spectral gap for the classical MCMC algorithm is added for comparison.}
    \label{fig:delta_vs_T}
\end{figure}

\begin{table*}[t]
\begin{tabular}{llll}
\hline\hline\noalign{\vskip 2pt}
   &  & Time per step and particle & Wallclock time for $N_\mathrm{Pa}\times T = 3\cdot 10^7$\\
\hline\noalign{\vskip 2pt}
   HSF $\qquad$  &SDE & $t_\mathrm{SDE} \approx 2.381\cdot 10^{-7} \mathrm{s}$ & $t_\mathrm{SDE} \approx 7.14\,\mathrm{s}$ \\
    &MCMC & $t_\mathrm{MCMC} \approx 3.465\cdot 10^{-7} \mathrm{s}$& $t_\mathrm{MCMC} \approx 10.39\,\mathrm{s}$ \\
    & QE-MCMC $\qquad$ & $t_\mathrm{QE-MCMC} \approx 1.699\cdot 10^{-6} \mathrm{s}\qquad$ & $t_\mathrm{QE-MCMC} \approx 50.96\,\mathrm{s}$\\
\\    
    & & Time per step and particle & Wallclock time for $N_\mathrm{Pa}\times T = 3\cdot 10^8$ \\
\hline\noalign{\vskip 2pt}
    TCF$\qquad$ &SDE & $t_\mathrm{SDE} \approx 5.607 \cdot 10^{-7} \mathrm{s}$ & $t_\mathrm{SDE} \approx 168\,\mathrm{s}$ \\
    &MCMC & $t_\mathrm{MCMC} \approx 1.770\cdot 10^{-6}\mathrm{s}$ & $t_\mathrm{MCMC} \approx 531\,\mathrm{s}$ \\
    &QE-MCMC & $t_\mathrm{QE-MCMC} \approx 7.490 \cdot 10^{-6} \mathrm{s} \qquad$ & $t_\mathrm{QE-MCMC} \approx 2247\,\mathrm{s}$ \\
\hline\hline
\end{tabular}
\caption{Wallclock time comparison of the three algorithms for the two cases, HSF and TCF. $N_{\rm Pa}$ is the number of particles and $T$ is the number of time steps. All comparisons are run on a single CPU.}
\label{tab:runtimes}
\end{table*}

The sampling temperature $T$ was chosen to 3.4 and 3.9 for the two- and one-dimensional sampling task, respectively. These are the same parameters which have been used for the calculation of the dispersion curves in Fig. \ref{fig:dispersion_curves_channel_flow}. Nevertheless, it is interesting to take a closer look at this free parameter in the present case, which does not stand for a real temperature as in the original canonical equilibrium ensemble in applications in condensed matter physics (cf. Eq. \eqref{eq:canonical}). The role of $T$ can be understood from two complementary perspectives.

From an algorithmic point of view, the higher the sampling temperature $T$, the more broadly the Markov chain will explore new accelerations. Smaller sampling temperatures cause a narrower target distribution and thus a more conservative selection with higher rejection rates, cf. Eq. \eqref{eq:acceptance}. This is clearly visible in Fig. \ref{fig:delta_vs_T} for $N_q>2$ and for both sampling tasks. If the temperature parameter is enhanced too strongly, however, the tracks of the velocity fluctuations will become too noisy. See also our discussion in subsection III.A.4.

From a representation point of view, in the QE-MCMC case, the acceleration space for each component is represented using only a few qubits, resulting in a coarse discretization of the possible acceleration values for the sampler. The nominal target may then be too sharply localized on this coarse grid and only a small number of discrete states receive a significant probability weight. The Markov chain has little room to explore the state space, in particular the tails of the distribution, and many quantum-generated proposals may land in states that are strongly penalized. This rationalizes why the tempered version works better.

Importantly, at the reference value $T=2$, the tempered target distribution reduces exactly to the original Gaussian distribution, so that QE-MCMC and classical MCMC operate under identical target statistics. This implies that any remaining differences in the spectral gap at $T=2$ are not caused by tempering, but reflect differences in the underlying transition mechanism of the Markov chain itself. In particular, for the bivariate sampling task the QE-MCMC already exhibits a larger spectral gap than the classical MCMC, whereas for the one-dimensional case the spectral gap of the QE-MCMC remains slightly smaller. At the same time, we still employ a tempered setting in this work, since it yields more robust dispersion statistics for the considered flow cases.

In summary, we could show that the quantum-enhanced MCMC algorithm yields a bigger effective spectral gap $\delta$ in case of the channel flow case with sampling from the most complex distribution in our applications, $p^{({\bf a})}_{\rm 2D}(a_x^+,a_y^+|y^+)$. Thus, our findings indicate an advantage of the quantum-enhanced case in comparison to the classical one for exactly those sampling problems that define potential future applications, namely complex distributions that depend on several variables.

\subsubsection{Runtime comparison}
To further assess the practical relevance of the proposed approach,  we compare their computational cost in terms of wall-clock time. Table~\ref{tab:runtimes} summarizes the runtimes of the SDE, classical MCMC, and QE-MCMC algorithms for both caes, HSF and TCF flow.

The results show that, on current classical hardware, the QE-MCMC algorithm is computationally more expensive than both, SDE and classical MCMC approaches. This is expected, as the quantum-enhanced sampling procedure involves additional overhead related to the circuit-based proposal generation and its emulation on classical processors. In contrast, the classical algorithms benefit from highly optimized implementations and efficient use of standard hardware architectures.

We note that the reported timings are obtained using optimized classical implementations based on just-in-time compilation \cite{lam2015}. In test cases without such optimization, the runtime differences between the methods become less pronounced, indicating that classical algorithms currently benefit more strongly from established hardware-specific acceleration techniques. This highlights that the present comparison is influenced by the maturity of classical computing frameworks, whereas comparable optimization strategies for quantum-enhanced algorithms are not yet available.

Thus, the runtime comparison must be interpreted with care. First, the present implementation does not yet benefit from actual quantum hardware acceleration, but relies on classical simulation of quantum circuits. Secondly, the advantage of QE-MCMC is not primarily reflected in raw wall-clock time, but rather in its sampling dynamics, as characterized by the spectral gap. In particular, for sampling tasks from more complex distributions, the QE-MCMC algorithm exhibits improved mixing properties. Note also that the ratio of wallclock times $t_{\rm QE-MCMC}/t_{\rm MCMC}=$ drops from 4.9 for the HSF case to 4.2 for the TCF case.

We therefore interpret the current findings as a proof of concept: while QE-MCMC is not yet competitive in terms of computational cost at the present technological stage, it already demonstrates favorable algorithmic properties for complex, strongly correlated sampling problems that are beyond the reach of simple factorized models. In this sense, the QE-MCMC approach should be viewed as complementary rather than competitive under current technological constraints.

\section{Summary and outlook}
The main objective of the present work was to apply a hybrid quantum-classical algorithm for the sampling of fluctuations of the acceleration components from well-resolved probability density functions that depend on several variables. These fluctuations are used to generate synthetic Lagrangian tracer tracks in two turbulent shear flow cases. These are (1) a homogeneous shear flow (HSF) -- a turbulent shear flow, which is characterized by a uniform shear rate -- and (2) a turbulent boundary layer flow, which was taken from the lower half-channel of a turbulent plane Poiseuille channel flow (TCF). In the second test case, data have to be sampled from a distribution that depends on more variables since the shear rate $S(y)$ is a function of the distance from the wall $y$; the shear rate is highest at the wall and decreases to zero at the center plane. The present work extends our previous study by Ingelmann et al. \cite{Ingelmann2025}. Both are to the best of our knowledge the first applications of quantum computing to Lagrangian fluid turbulence (which considers the fluid motion in a frame of reference that is co-moving with fluid parcels). 

The statistical analysis of the synthetic Lagrangian tracer tracks was focused on the scaling of particle pair dispersion with respect to time. This is a central quantity that characterizes the turbulent mixing properties in a turbulent flow as the prefactor in the corresponding law quantifies the scale-dependent turbulent diffusivity. Since turbulent shear flows are anisotropic, these dispersions scale differently with respect to different directions in space. For short times, all algorithms give a ballistic scaling, $D(t)\sim t^2$. For the largest times, we obtain a diffusive scaling, $D_i(t)\sim t$. At intermediate time scales of the order of the shear time scale, we find a superdiffusive scaling of the mean pair dispersion, $D(t)\sim t^{\alpha}$ with exponents $\alpha \simeq 3$ to 4. The main reason for the latter scaling are the streamwise streaky structures of size $\ell_{\rm corr}$, which we implemented in the stochastic model via the correlation decay term in the otherwise featureless synthetic fields. To summarize, the introduction of decaying spatial correlations and two time scales, $\tau_i$ and $\tau_{\eta}<\tau_i$, are thus already capable to generate realistic Lagrangian turbulent dispersion.

The two-layer stochastic model is based on the velocity fluctuation statistics, which we obtained in both cases from simulation data. Velocity fluctuation statistics is used to infer acceleration statistics, which is required for the stochastic equation system. The latter is thus also Gaussian in the present stochastic approach. It is known that accelerations are typically non-Gaussian distributed \cite{Toschi2009}. This would require a extension of the present two-layer stochastic model to multiple layers (and thus regressive equations), thus introducing a whole hierarchy of time scales of the turbulent flow, as discussed in Viggiano et al. \cite{Viggiano2020}. Such a more realistic model goes beyond the scope of the present proof-of-concept study, but will be investigated further in the future.

However, we could demonstrate the applicability of the hybrid quantum-classical generative algorithm for a use case from classical fluid dynamics. The present quantum sampling algorithm requires a one-shot measurement only that keeps the query complexity in applications low. A further positive aspect is that the method operates already robustly with a low number of qubits (here $N_q=5$ or 6) and is thus appropriate for the present noisy intermediate scale quantum devices.

A significantly larger effective spectral gap in QE-MCMC than MCMC for the bivariate PDF of the more complex TCF case suggests an advantage of the quantum algorithm in comparison to its classical counterpart. It is seen for the TCF tempered sampling cases with best resolution due to largest qubit number, see Fig. \ref{fig:delta_vs_N1}. In contrast, wall-clock measurements indicate that the current QE-MCMC implementation is computationally more expensive than the classical MCMC approach, which reflects the additional overhead of the circuit-based proposal generation and its classical emulation rather than an intrinsic limitation of the sampling strategy itself.

If this finding is robust for other flow cases needs to be explored in future investigations. This would encourage further QE-MCMC applications in more complex flow configurations. These would generate joint distributions that depend on more physical variables and parameters, i.e., involve more degrees of freedom. An example can be the particle transport in rotating spherical-shell and buoyancy-driven turbulent flows. These investigations are in progress and will be reported elsewhere. 

It also remains to be clarified whether a quantum computing module, such as the present Lagrangian one which processes a smaller subtask for many tracers, can be integrated into a larger classical high-performance computing framework for turbulent fluid flow. In particular, the potential of quantum-centric supercomputing platforms, where a quantum processor is coupled to a classical HPC system, for modeling turbulent transport of pollutants, aerosols, or chemical species in a hybrid Earth system model remains an open question. The present work represents, in our view, a first promising step in this direction.

\section*{Acknowledgements}
The work is funded by the European Union (ERC, MesoComp, 101052786). Views and opinions expressed, however, are those of the authors only and do not necessarily reflect those of the European Union or the European Research Council. F.S. is supported by the Deutsche Forschungsgemeinschaft (DFG). The work benefited from helpful discussions with Sachin S. Bharadwaj and Katepalli R. Sreenivasan.

\section*{Data Availability}
The input data for the homogeneous shear flow case are found in Zenodo \cite{Zenodo}. The input data for the turbulent channel case have been taken from the Johns Hopkins Turbulence Database (JHTDB) \cite{Li2008}. These are the channel flow data at $Re_{\tau}=1000$ \cite{Graham2016}. The Python scripts for processing the data and the MCMC and QE-MCMC algorithms in this manuscript are found in a further repository in Zenodo.

\appendix
\section{Diffusion matrix $\hat G$ from velocity fluctuation covariance 
$\hat \Sigma_{{\bf u}'{\bf u}'}$}

In this appendix, we explain in brief how the diffusion matrix $\hat G$ in the SDE for the accelaration components is inferred in our approach from the velocity fluctuation covariance $\hat \Sigma_{{\bf u}'{\bf u}'}$, the latter of which can be extracted from the simulation data bases, see also ref. \cite{Sarkka2019} for details. We use the compact block matrix notation for the following and summarize Eqs. \eqref{SDE1a} and \eqref{SDE1b} to
\begin{equation}
\mathrm d{\bf z}=\hat A\, {\bf z}\, \mathrm dt + \hat B\,\mathrm d{\bf W} 
\end{equation}
with
\begin{align} 
{\bf z} = \left(\begin{array}{l} {\bf a}  \\ {\bf u}^{\prime} \end{array} \right), \; \hat A = \left(\begin{array}{rr} -\hat T & \hat I \\ \hat I  & -\dfrac{1}{\tau_{\eta}}\hat I \end{array}\right), \;
\hat B = \left(\begin{array}{l} 0  \\ \hat{G} \end{array} \right),
\end{align}
containing identity matrix $\hat I$ and matrix $\hat T$ with the autocorrelation times of the three components of the velocity fluctuation vector field, $\hat T={\rm diag}(\tau_1,\tau_2,\tau_3)$. Next, we define the covariance block matrix for $\bf z$ as follows,
\begin{align} 
\hat \Sigma_{{\bf z}{\bf z}} = \left(\begin{array}{cc} \hat\Sigma_{{\bf u}'{\bf u}'} & \hat\Sigma_{{\bf u}'{\bf a}} \\ \hat\Sigma_{{\bf a}{\bf u}'} & \hat\Sigma_{{\bf a} {\bf a}}\end{array}\right) \in \mathbb{R}^{6\times 6}\,. 
\end{align}
Here, $\hat\Sigma=\hat\Sigma_{{\bf u}'{\bf u}'}$ to distinguish it better from the remaining matrices. The covariance matrix $\hat \Sigma_{{\bf z}{\bf z}}$ satisfies the Lyapunov equation \cite{Sarkka2019}, which is given by
\begin{align} 
\frac{{\rm d}\hat \Sigma_{{\bf z}{\bf z}}}{{\rm d}t} = \hat A \hat \Sigma_{{\bf z}{\bf z}} + \hat \Sigma_{{\bf z}{\bf z}} \hat A^\mathrm{T} + \hat B \hat B^\mathrm{T}\,. 
\label{Lyapunov}
\end{align}
We seek statistically steady distributions, such that the left hand side of \eqref{Lyapunov} is zero. In detail, we get with $\hat\Sigma_{{\bf a}{\bf u}'} =\hat\Sigma_{{\bf u}'{\bf a}}$ the following terms,
\begin{align*} 
\hat A\hat \Sigma_{{\bf z}{\bf z}} &= \left(\begin{array}{cc} 
-\hat{T}^{-1}\hat\Sigma_{{\bf u}'{\bf u}'} + \hat\Sigma_{{\bf u}'{\bf a}'} & 
-\hat{T}^{-1}\hat\Sigma_{{\bf u}'{\bf a}} + \hat\Sigma_{{\bf a}{\bf a}}\\
-\frac{1}{\tau_{\eta}}\hat\Sigma_{{\bf u}'{\bf a}} & 
-\frac{1}{\tau_{\eta}}\hat\Sigma_{{\bf a} {\bf a}}\end{array}\right) \\
\hat \Sigma_{{\bf z}{\bf z}} \hat A^\mathrm{T}&= \left(\begin{array}{cc} 
-\hat\Sigma_{{\bf u}'{\bf u}'}\hat T^{-1} + \hat\Sigma_{{\bf u}'{\bf a}} & 
-\frac{1}{\tau_{\eta}}\hat\Sigma_{{\bf u}'{\bf a}} \\
-\hat\Sigma_{{\bf u}'{\bf a}} \hat{T}^{-1} + \hat\Sigma_{{\bf a}{\bf a}} &
-\frac{1}{\tau_{\eta}}\hat\Sigma_{{\bf a} {\bf a}}\end{array}\right) \\
\hat B\hat B^\mathrm{T} &= \left(\begin{array}{cc}
0 & 0 \\ 0 & \hat G \hat G^\mathrm{T}\end{array}\right)\,. 
\end{align*}
We can solve the steady Lyapunov equation blockwise:
\begin{align} 
{\rm tl} &: \quad  \hat\Sigma_{{\bf u}'{\bf a}}=\frac{1}{2}(
\hat{T}^{-1}\hat\Sigma_{{\bf u}'{\bf u}'} + \hat\Sigma_{{\bf u}'{\bf u}'}\hat T^{-1})\\
{\rm tr} & : \quad \hat\Sigma_{{\bf a}{\bf a}}=\left(
\hat{T}^{-1} + \frac{1}{\tau_{\eta}} \hat I\right) \hat\Sigma_{{\bf u}'{\bf a}} \\
{\rm bl} & : \quad \hat\Sigma_{{\bf a}{\bf a}}=\hat\Sigma_{{\bf u}'{\bf a}}\left(
\hat{T}^{-1} + \frac{1}{\tau_{\eta}} \hat I\right) \\
{\rm br} & : \quad \hat G \hat G^\mathrm{T}= \frac{2}{\tau_{\eta}}\Sigma_{{\bf a}{\bf a}}\,, 
\end{align}
where tl stands for top left, tr for top right, bl for bottom left, and br for bottom right, respectively. Combining the conditions bl and tl in br, one ends up with 
\begin{align} 
\hat G\hat G^\mathrm{T}=\frac{1}{\tau_{\eta}}\left(
\hat{T}^{-1} + \frac{1}{\tau_{\eta}} \hat I\right)(
\hat{T}^{-1}\hat\Sigma_{{\bf u}'{\bf u}'} + \hat\Sigma_{{\bf u}'{\bf u}'}\hat T^{-1})\,, 
\end{align}
which connects $\hat G$ and $\hat\Sigma_{{\bf u}'{\bf u}'}$ and allows to calculate the components of the diffusion matrix $\hat G$.

\bibliography{bibliography}

\end{document}